\documentclass[twocolumn]{aastex631}

\usepackage{amsmath}

\def\aap{A\&A}

\def\apjl{ApJL}

\def\apjs{ApJS}

\begin{document}

\title{Potential Contribution of Young Pulsar Wind Nebulae to Galactic High-Energy Neutrino Emission}

\author[0000-0003-3089-3762]{Xuan-Han Liang}
\affiliation{School of Astronomy and Space Science, Nanjing University, 163 Xianlin Avenue,\\
Nanjing 210023, Jiangsu, People's Republic of China}
\affiliation{Key Laboratory of Modern Astronomy and Astrophysics, Nanjing University, Ministry of Education,\\
Nanjing 210023, Jiangsu, People's Republic of China}

\author[0000-0003-4907-6666]{Xiao-Bin Chen}
\affiliation{School of Astronomy and Space Science, Nanjing University, 163 Xianlin Avenue,\\
Nanjing 210023, Jiangsu, People's Republic of China}
\affiliation{Key Laboratory of Modern Astronomy and Astrophysics, Nanjing University, Ministry of Education,\\
Nanjing 210023, Jiangsu, People's Republic of China}

\author{Ben Li}
\affiliation{Gran Sasso Science Institute, Viale F. Crispi 7 -- I-67100 L'Aquila, Italy}
\affiliation{INFN-Laboratori Nazionali del Gran Sasso, Via G. Acitelli 22, 67100 Assergi (AQ), Italy}

\author[0000-0003-1576-0961]{Ruo-Yu Liu}
\affiliation{School of Astronomy and Space Science, Nanjing University, 163 Xianlin Avenue,\\
Nanjing 210023, Jiangsu, People's Republic of China}
\affiliation{Key Laboratory of Modern Astronomy and Astrophysics, Nanjing University, Ministry of Education,\\
Nanjing 210023, Jiangsu, People's Republic of China}
\affiliation{Tianfu Cosmic Ray Research Center,\\ Chengdu 610000, Sichuan, People's Republic of China}
\correspondingauthor{Ruo-Yu Liu}
\email{ryliu@nju.edu.cn}

\author[0000-0002-5881-335X]{Xiang-Yu Wang}
\affiliation{School of Astronomy and Space Science, Nanjing University, 163 Xianlin Avenue,\\
Nanjing 210023, Jiangsu, People's Republic of China}
\affiliation{Key Laboratory of Modern Astronomy and Astrophysics, Nanjing University, Ministry of Education,\\
Nanjing 210023, Jiangsu, People's Republic of China}

\begin{abstract}

Pulsar wind nebulae (PWNe), especially the young ones, are among the most energetic astrophysical sources in the Galaxy. It is usually believed that the spin-down energy injected from the pulsars is converted into magnetic field and relativistic electrons, but the possible presence of proton acceleration inside PWNe cannot be ruled out. Previous works have estimated the neutrino emission from PWNe using various source catalogs measured in gamma-rays. However, such results rely on the sensitivity of TeV gamma-ray observations and may omit the contribution by unresolved sources. Here we estimate the potential neutrino emission from a synthetic population of PWNe in the Galaxy with a focus on the ones that are still in the free expansion phase. In the calculation, we model the temporal evolution of the free-expanding PWNe and consider the transport of protons inside the PWNe. The Crab nebula is treated as a standard template for young PWNe to evaluate some model parameters, such as the energy conversion fraction of relativistic protons and the target gas density for the hadronic process, which are relevant to neutrino production. In the optimistic case, the neutrino flux from the simulated young PWNe may constitute to 5\% of the measured flux by IceCube around 100~TeV. At higher energy around 1~PeV, the neutrino emission from the population highly depends on the injection spectral shape, and also on the emission of the nearby prominent sources.

\end{abstract}


\section{Introduction} \label{sec: intro}

Cosmic rays (CRs) are charged particles wandering in the universe. When CRs interact with gas, gamma-rays and neutrinos are produced through neutral pion and charged pion decay, respectively. On the other hand, inverse Compton (IC) scattering process of electron/positron pairs (hereafter we use electrons to represent both of them for simplicity) against the radiation field can also produce gamma-rays. Hence, neutrino emission is regarded as the smoking gun of hadronic processes and acceleration of CR hadrons. Unlike CRs that will be deviated by the magnetic field, neutrinos hardly interact with matter and carry the unique information of the interactions directly to Earth. Great efforts have been put into gigantic detector construction in ice or water all over the world to capture high-energy neutrinos, including IceCube \citep{Aartsen2017JInst..12P3012A}, ANTARES \citep{Ageron2011NIMPA.656...11A}, KM3NeT \citep{Adrian-Martinez2016JPhG...43h4001A}, and Baikal-GVD \citep{Avrorin2011NIMPA.639...30A}, etc.

The Milky Way has been predicted to be a source of high-energy neutrinos \citep{Stecker1979ApJ...228..919S,Evoli2007JCAP...06..003E}, yet analyses of earlier observations found no significant excess \citep{Adrian-Martinez2016PhLB..760..143A,Albert2017PhRvD..96f2001A,Aartsen2017ApJ...849...67A,Albert2018ApJ...868L..20A}. An independent study using track events from IceCube above 200~TeV showed $|b|_{\text{med}}\approx21^{\circ}$, where $b$ is the Galactic latitude and "med" stands for median, indicating a deviation towards lower latitude from isotropy implied by the null hypothesis and thus a component of high-energy neutrino flux of Galactic origin \citep{Kovalev2022ApJ...940L..41K}. Recently, using cascade events from IceCube, the high-energy neutrino emission was identified from the Galactic plane at the $4.5\sigma$ level of significance by comparing three diffuse emission models, which were referred to as $\pi^0$, $\text{KRA}_{\gamma}^{5}$, and $\text{KRA}_{\gamma}^{50}$, to a background-only hypothesis \citep{Abbasi2023Sci...380.1338I}. Nevertheless, there was no statistical evidence strong enough to differentiate between the three models. On the other hand, catalog stacking analyses of supernova remnants (SNR), pulsar wind nebulae (PWN), and other unidentified (UNID) Galactic sources showed $3\sigma$ excess for all three types. In other words, point sources may contribute to the emission at an unknown level.

Filled with relativistic particles, a PWN is a bubble formed by the interaction between the pulsar's relativistic wind and its environment \citep{Gaensler2006ARA&A..44...17G}. The spin-down energy of the energetic pulsar is believed to deposit into magnetic field and relativistic electrons, but protons or heavier nuclei may also exist. Although the IC scattering process of electrons is widely accepted as the dominant process of gamma-ray production in the PWN, part of the emission may be related to protons \citep{Cheng1990JPhG...16.1115C,Atoyan1996MNRAS.278..525A,Amato2003A&A...402..827A,Horns2006A&A...451L..51H,Zhang2009ApJ...699L.153Z,Zhang2020MNRAS.497.3477Z,Peng2022ApJ...926....7P,Nie2022ApJ...924...42N,Liang2022Univ....8..547L,Chen2024ApJ...976..172C}. Therefore, fitting the multi-wavelength energy spectrum can help determine the proton energy fraction. On the other hand, PWNe have been suggested to be possible Galactic neutrino sources (e.g. \citealt{Bednarek2003A&A...407....1B}). Using different source catalogs, previous works have tried to estimate the neutrino emission from PWNe and the corresponding neutrino events for different detectors (e.g. \citealt{Guetta2003APh....19..403G,DiPalma2017ApJ...836..159D,Gagliardini2024ApJ...969..161G}). Some sources, e.g. Crab, Vela-X, were found to be point source candidates. Apart from focusing on individual PWNe, catalog stacking analysis has also been conducted. \citet{Gagliardini2024ApJ...969..161G} claimed that stacking predicted KM3NeT data accumulated in 1 year from $\sim3$ promising PWNe can reach a significance value of about $7.5\sigma$, but 9.5-year all-sky IceCube data from 35 PWNe were found to be absent of a significant excess \citep{Aartsen2020ApJ...898..117A}. However, either individual estimation or stacking analysis is limited to the size of the catalog and thus is intrinsically biased.

To get the panorama of the possible neutrino emission from the PWNe, we do not follow the conventional methods but instead calculate the emission from a synthetic PWN population directly and evaluate its contribution. In this work, we focus on the young PWNe in the Galaxy, which are still in the free expansion phase. The Crab nebula is used as the standard template for the population study in order to get the proton energy fraction, as well as to estimate the amount of target gas in the nebula.

The remainder of the paper is structured as follows. A dynamic PWN model is described in Section \ref{sec: method} and is applied to the Crab nebula in Section \ref{sec: Crab}. In Section \ref{sec: population} we generate the young PWN sample and calculate the neutrino emission. The result is discussed in Section \ref{sec: dis} and the conclusion is given in Section \ref{sec: con}.

\section{PWN Model description} \label{sec: method}

\subsection{Evolution of PWN in the free expansion phase}

Generally speaking, there are three phases in the evolution of a PWN (e.g. \citealt{Gaensler2006ARA&A..44...17G}). Before the reverse shock of the SNR collides with the outer boundary of the PWN, the PWN interacts with the freely expanding unshocked SNR ejecta, and for this reason, this stage is called the free expansion phase. After the collision, the reverberation phase begins, during which the PWN is firstly compressed and then re-expands due to the difference of pressure inside and outside the boundary. This phase can be rather complex, and the compression-expansion process may even repeat several times with damping \citep{Bandiera2023MNRAS.520.2451B}. As the reverberation gradually fades, the PWN enters the Sedov-Taylor phase. Here, we only focus on the PWN that is still in its early evolution phase. There are several reasons for our choice: (i) the central pulsar is young and thus the energy budget for the proton injection from the spin-down process is more abundant than that in the later phase; (ii) the polarized observation of \textit{IXPE} towards the Crab nebula suggests a predominantly toroidal magnetic field \citep{Bucciantini2023NatAs...7..602B}, implying that in this pre-reverberation stage, advection is likely to be dominant for particle transport, and consequently, particles, including the most energetic ones, will not escape easily, leading to relatively high neutrino production inside the system; (iii) simple dynamic models of spherical symmetry are robust to describe this phase \citep{Gelfand2009ApJ...703.2051G,Tanaka2010ApJ...715.1248T,Bucciantini2011MNRAS.410..381B,Martin2012MNRAS.427..415M,Vorster2013ApJ...773..139V,Torres2013MNRAS.436.3112T,Torres2014JHEAp...1...31T,Lu2017ApJ...834...43L,Martin2022JHEAp..36..128M}, while it still requires further effort to fully understand the later stages.

First of all, we need to trace the evolution of the PWN. Following \citet{Truelove1999ApJS..120..299T}, the characteristic radius and time of the relevant SNR are given by
\begin{equation}
    R_{\text{ch}}=M_{\text{ej}}^{1/3}\rho_{\text{ism}}^{-1/3},
\end{equation}
\begin{equation}
    t_{\text{ch}}=M_{\text{ej}}^{5/6}E_{\text{sn}}^{-1/2}\rho_{\text{ism}}^{-1/3},
\end{equation}
where $M_{\text{ej}}$ is the mass of the ejecta, $E_{\text{sn}}$ is the energy of the supernova explosion and $\rho_{\text{ism}}=1.4m_{p}n_{\text{ism}}$ is the mass density of the interstellar medium (ISM). \citet{Bandiera2023MNRAS.520.2451B} considered a density distribution in the SNR ejecta: $\rho_{\mathrm{ej}}(t)=\mathcal{A}/t^{3}$, where $\mathcal{A}=5E_{\mathrm{sn}}/(2\pi)\left[3M_{\mathrm{ej}}/\left(10E_{\mathrm{sn}}\right)\right]^{5/2}$. The approximations of the outer boundary of the PWN, $R_{\text{pwn}}$, and of the reverse shock, $R_{\text{rs}}$, are then derived as
\begin{equation}
    \begin{aligned}
        R_{\text{pwn}}(t)=& \mathcal{V}_0\tau_0 \left[ 1+\left( C_{R,0}^{5/6}\frac{t}{\tau_0} \right)^{-a} \right]^{-6/(5a)}\\
        &\times\left[ 1+\left( C_{R,\infty}\frac{t}{\tau_0} \right)^{b} \right]^{1/b}
    \end{aligned},
    \label{eq: R_pwn}
\end{equation}
\begin{equation}
    R_{\text{rs}}(t)=\frac{12.49\,(2.411-t/t_{\text{ch}})^{0.6708}\,(t/t_{\text{ch}})^{1.663}}{1+17.47\,t/t_{\text{ch}}+4.918\,(t/t_{\text{ch}})^2}R_{\text{ch}}, \label{eq: R_rs}
\end{equation}
where $\mathcal{V}_0=\left(L_{0}\tau_{0}/\mathcal{A}\right)^{1/5}$. The formula of $R_{\text{pwn}}$ has 4 parameters and among them $C_{R,0}\simeq 0.7868$. The rest are fitted using third-degree polynomials with the braking index $n$ being the variable written as $C_{R,\infty}(n) \simeq 0.30139 + 0.46268 n - 0.099087 n^2 + 0.008715 n^3$, $a(n) \simeq 0.89882 - 0.00365 n - 0.045432 n^2 + 0.006836 n^3$, and $b(n) \simeq 0.78755 + 0.08107 n - 0.068173 n^2 + 0.008969 n^3$. More details can be found in the Appendix D in \citet{Bandiera2023MNRAS.520.2451B}. We use their result for both the Crab nebula and the synthetic sample.

On the other hand, the radius of the termination shock, $R_{\text{ts}}$, is estimated by
\begin{equation}
    R_{\text{ts}}(t)=\sqrt{\frac{L(t)}{4 \pi c P_{\text{pwn}}(t)}}. \label{eq: r_ts}
\end{equation}
The spin-down process of the central pulsar follows
\begin{equation}
    L(t) = L_0 \left(1 + \frac{t}{\tau_0}\right)^{-\frac{n+1}{n-1}}, \label{eq: spin-down L}
\end{equation}
where $L_0$ is the initial spin-down luminosity, $\tau_0$ is the initial spin-down timescale, and $n$ is the braking index. The pressure of the PWN is computed as
\begin{equation}
    P(t)=\frac{1}{4 \pi R^{4}(t)} \int_{0}^{t} \text{d}t^{\prime} \, L\left(t^{\prime}\right) R\left(t^{\prime}\right).
\end{equation}
Once $R_{\text{ts}}$ calculated by Eq. \ref{eq: r_ts} reaches 0.13 pc for the Crab \citep{Weisskopf2012ApJ...746...41W}, we fix this value in order to match the observation. For the synthetic sample, we choose 0.1 pc as the general case.

The structure of the flow and the magnetic field in PWNe is nontrivial as shown by magnetohydrodynamic (MHD) simulations (e.g. \citealt{DelZanna2004A&A...421.1063D,Porth2014MNRAS.438..278P}) as well as a recent 3D mapping of the Crab \citep{Martin2021MNRAS.502.1864M} . Here in the spherical model a radial flow and azimuthal magnetic field is assumed along with the ideal MHD limit $\nabla \times \boldsymbol{V} \times \boldsymbol{B}=0$, which then reduces to $VBr={\rm constant}=V_0B_0r_0$ with the subscript "0" representing the value at the termination shock \citep{Vorster2013ApJ...765...30V}. According to \citet{Kennel1984ApJ...283..694K}, the velocity profile depends on the magnetization parameter $\sigma$, and the velocity decreases with increased radius. For the current model, the bulk velocity is expressed as
\begin{equation}
    V(r,t) = V_f\frac{R_{\text{pwn}}(t)}{t}\left( \frac{r}{R_{\text{pwn}}(t)} \right)^{-\beta}, \label{v}
\end{equation}
where $0 \le V_f \le 1$. Correspondingly, the radial profile of the magnetic field inside the PWN is then given by
\begin{equation}
    B(r,t)=B_{0}(t)\left(\frac{r}{R_{\text{ts}}(t)}\right)^{\beta-1}. \label{eq: B}
\end{equation}
The magnetic field at the termination shock, $B_{0}$, can be obtained by solving
\begin{equation}
    \frac{\text{d}W_{\text{B}} (t)}{\text{d}t}=\eta_{\text{B}}L(t)-\frac{W_{\mathrm{B}}(t)}{R_{\text{pwn}}(t) }\frac{\text{d}R_{\text{pwn}}(t)}{\text{d}t},
\end{equation}
where $W_{\text{B}}(t)=\int \text{d}r\,r^2B^{2}(r,t)/2$, and $\eta_{\text{B}}$ is the ratio of the magnetic energy to the spin-down energy \citep{Torres2014JHEAp...1...31T}. Recent phenomenological analyses of the Crab's spectrum with the magnetic field profile similar to Eq.~(\ref{eq: B}) found $\beta\approx0.5$ \citep{Dirson2023A&A...671A..67D,Aharonian2024A&A...686A.308A}, which was also the value used by \citet{Vorster2013ApJ...765...30V}. Hereafter, we fix $\beta=0.5$.

\subsection{Particles in the PWN}

In this work, we use the model proposed by \citet{Vorster2013ApJ...765...30V} and subsequently employed by \citet{Lu2017ApJ...834...43L} and \citet{Peng2022ApJ...926....7P}. In the spherically symmetric model, the number density of particles $n=n(r,\gamma,t)$ within the PWN can be obtained by solving the transport equation (e.g. \citealt{Parker1965P&SS...13....9P})
\begin{equation}
    \begin{aligned}
        \frac{\partial n}{\partial t}=& D\frac{\partial^2n}{\partial r^2}+\left[\frac{1}{r^2}\frac{\partial}{\partial r}(r^2D)-V\right]\frac{\partial n}{\partial r}\\
        &-\frac{1}{r^2}\frac{\partial}{\partial r}[r^2V]n+\frac{\partial}{\partial\gamma}[\dot{\gamma}n]+Q_{\text{inj}}, \label{eq: distribution}
    \end{aligned}
\end{equation}
where $V$ is the flow velocity given by Eq. \ref{v}, $D$ is the diffusion coefficient, $\dot{\gamma}$ is the summation of energy losses, and $Q_{\text{inj}}$ is the injection term. The spin-down energy of the pulsar is converted into magnetic field and relativistic particles. In this work, both electrons and protons are taken into account, i.e. $\eta_{\text{B}}+\eta_{\text{e}}+\eta_{\text{p}}=1$. The energy fractions are free parameters in the calculation and can be determined once two of them are given. It is assumed that the particles are injected at the termination shock, with the injection rate being
\begin{equation}
    Q_{\mathrm{inj}}^{\mathrm{e}}(\gamma_{\mathrm{e}},t)=Q_{0}^{\mathrm{e}}(t)
    \left\{
    \begin{array}{l}
        \left(\frac{\gamma_{\mathrm{e}}}{\gamma_{\mathrm{b}}}\right)^{-\alpha_{1}} \quad \gamma_{\mathrm{e}, \min} \leq \gamma_{\mathrm{e}} < \gamma_{\mathrm{b}} \\
        \left(\frac{\gamma_{\mathrm{e}}}{\gamma_{\mathrm{b}}}\right)^{-\alpha_{2}} \quad \gamma_{\mathrm{b}} \leq \gamma_{\mathrm{e}} \leq \gamma_{\mathrm{e}, \max}
    \end{array}
    \right.
\end{equation}
for electrons and
\begin{equation}
    Q_{\text{inj}}^{\text{p}}\left(\gamma_\text{p},t\right) =Q_{0}^{\text{p}}(t)\gamma_{\text{p}}^{-\alpha_{\text{p}}}e^{-\gamma_{\text{p}}/\gamma_{\text{p,c}}}
\end{equation}
for protons, respectively. The normalization terms $Q_{0}$ can be obtained through $\eta L(t)=\int \text{d}\gamma\,\gamma mc^2Q_{\text{inj}}(\gamma,t)$.

The maximum energy achieved by acceleration within the termination shock changes with the spin-down history and is estimated by
\begin{equation}
    E_{\max}(t)=\varepsilon e \kappa \sqrt{\frac{\eta_{\text{B}} L(t)}{c}},
\end{equation}
where $0 < \varepsilon \leq 1$ is the ratio between the Larmor radius of the relativistic particle $r_{\text{L}}=E/(ZeB)$ and $R_{\text{ts}}$, and $\kappa=3$ is the magnetic compression ratio \citep{Martin2022JHEAp..36..128M}. Hereafter we fix $\varepsilon=1$ to consider the optimistic case. Considering the typical Kolmogorov turbulence with energy dependence $\propto E^{1/3}$ and the spacial variance related to the magnetic field $\propto 1/B(r,t)$, the diffusion coefficient can be expressed as
\begin{equation}
    D\left(r,E,t\right)=D_{0}\left(\frac{r}{R_{\text{ts}}(t)}\right)^{1-\beta}\left(\frac{E}{E_{\max}}\right)^{\frac{1}{3}},
\end{equation}
where $D_0=cE_{\max}/(3eB_0)$ is the diffusion coefficient for the maximum energy at the termination shock.

Due to the expansion of the PWN, both electrons and protons suffer from the adiabatic loss:
\begin{equation}
    \dot{\gamma}_{\text{ad}}(r,\gamma,t)=-\frac{1}{3}\nabla\cdot \boldsymbol{V} \gamma=-\frac{1}{3}\left[\frac{2V}{r}+\frac{\partial V}{\partial r}\right]\gamma.
\end{equation}
The magnetic field inside the Crab nebula has been shown by both MHD simulations (e.g. \citealt{Porth2014MNRAS.438..278P,Olmi2016JPlPh..82f6301O}) and spectral analyses (e.g. \citealt{Peng2022ApJ...926....7P,Dirson2023A&A...671A..67D,Aharonian2024A&A...686A.308A}) to be around several hundred microgauss in recent researches. Therefore, the electrons are subject to severe synchrotron loss given by
\begin{equation}
    \dot{\gamma}_{\mathrm{syn}}(r,\gamma_{\mathrm{e}},t)=-\frac{4}{3}\frac{\sigma_{\mathrm{T}}}{m_{\mathrm{e}}c}\gamma_{\mathrm{e}}^{2}U_{\mathrm{B}}(r,t).
\end{equation}

To equate the amount of particles that are injected to the amount of particles that flows into the nebula, the following inner boundary condition should be satisfied:
\begin{equation}
    V_0n-D(R_{\text{ts}},\gamma,t)\frac{\partial n}{\partial r}=\frac{Q_{\text{inj}}(\gamma,t)}{4\pi R_{\text{ts}}^2(t)},
\end{equation}
where $V_0$ is the velocity at the termination shock. On the other hand, the free-escape condition is imposed at the outer boundary of the PWN, i.e. $n(R_{\text{pwn}},\gamma,t)=0$ \citep{Vorster2013ApJ...765...30V}.

\subsection{Radiation process}

The synchrotron power per unit frequency emitted by a single electron is written as
\begin{equation}
    P_{\text{syn}}(r,\nu,\gamma,t)=\frac{\sqrt{3}e^{3}B(r,t)}{m_{e}c^{2}}F\left(\frac{\nu}{\nu_{c}}\right)
\end{equation}
where $\nu_{c}=3\gamma^2eB/4\pi m_e c$ is the critical frequency. We use an approximation of $F(\nu/\nu_{c})$ from \citet{Fouka2013RAA....13..680F}. The isotropic synchrotron emissivity (in $4\pi$ solid angle) is then expressed as
\begin{equation}
    \begin{aligned}
        Q_{\text{syn}}(r,\nu,t)&=4\pi j_{\text{syn}}(r,\nu,t)\\
        &=\int_0^\infty \text{d}\gamma \, n_e(r,\gamma,t)P_{\text{syn}}(r,\nu,\gamma,t).
    \end{aligned}
\end{equation}
Correspondingly, the power as well as the isotropic emissivity of IC scattering process from a single electron are respectively given by
\begin{equation}
    P_{\text{IC}}(r,\nu,\gamma,t)=\frac{ 2\pi r_0^2 c}{\gamma^2} \int_0^\infty \frac{n_{\mathrm{seed}}(r,\epsilon)\text{d}\epsilon}{\epsilon}f_{\text{IC}}(\epsilon,\nu,\gamma),
\end{equation}
\begin{equation}
    \begin{aligned}
        Q_{\text{IC}}(r,\nu,t)&=4\pi j_{\text{IC}}(r,\nu,t)\\
        &=\int_0^\infty \text{d}\gamma \, n_e(r,\gamma,t)P_{\text{IC}}(r,\nu,\gamma,t).
    \end{aligned}
\end{equation}
$f_{\text{IC}}$ is the general function for scattering of electrons in an isotropic photon gas \citep{Jones1968PhRv..167.1159J}. Following \citet{Dirson2023A&A...671A..67D}, three main seed photon fields contributing to the IC scattering in the Crab Nebula are taken into account: (i) the 2.73K cosmic microwave background radiation (CMB); (ii) the synchrotron radiation; (iii) the FIR radiation associated to the dust emission, which is described in their Section 3.3. The spectral number density $\text{d}n = n_{\text{seed}}(r,\epsilon)\text{d}V\text{d}\epsilon$ ($\epsilon=h\nu$) of seed photons from (ii) and (iii) are inhomogeneous in the nebula, whose value at a distance r to the center of the nebula in the spherical configuration is determined by \citep{Atoyan1989A&A...219...53A}
\begin{equation}
    n_{\mathrm{seed}}(r,\epsilon)=\frac{4 \pi}{h \epsilon} \frac{1}{2 c} \int_{r_{\min}}^{r_{\max}}\text{d} r_1 \,\frac{r_1}{r} j_v(r_1, \epsilon) \ln \left(\frac{r+r_1}{|r-r_1|}\right). \label{eq: photon seed}
\end{equation}

As for the hadronic process, we use a \textsc{python} package \texttt{aafragpy} \citep{Koldobskiy2021PhRvD.104l3027K} to obtain the differential cross section in the proton-proton interaction. The isotropic emissivity is given by
\begin{equation}
    Q_{\text{s}}(r,E_{\text{s}},t)=cn_{\text{gas}}\int_{4\,\text{GeV}}^{\infty}\mathrm{d}E_{\text{p}}\,\sigma_{\text{s}}(E_{\text{p}},E_{\text{s}})n_{\text{p}}(r,E_{\text{p}},t),
\end{equation}
where $n_{\text{gas}}$ is the gas density, and $s=\gamma,\nu$ denotes the secondary gamma-ray and neutrino, respectively. The lower limit of 4 GeV is a constraint due to the package, but it will not affect the result as we are interested in the TeV-PeV regime.

Finally, assuming spherical symmetry and optically thin plasma, the flux of different kinds of emission can all be calculated by
\begin{equation}
    E^2 \frac{dN}{dE} = \frac{E^2}{4 \pi d^2} \int_{R_{\text{ts}}}^{R_{\text{pwn}}} \text{d}r \, 4\pi r^2 Q(r,E),
\end{equation}
where $d$ is the distance of the source.

\section{Spectrum Fitting of the Crab nebula} \label{sec: Crab}

The Crab nebula is associated to a supernova explosion recorded by the ancient Chinese astrologer in 1054 AD and may be one of the most famous and well-studied sources in astrophysics \citep{Hester2008ARA&A..46..127H}. Numerous observations towards this astrophysical laboratory have already covered the electromagnetic spectrum from radio to ultra-high-energy (UHE) gamma-ray. Recently, PeV-photon detection by LHAASO made it a confirmed PeVatron in the Galaxy \citep{Cao2021Sci...373..425L}. In the following we first present the fitting result of its multi-wavelength spectrum, and then focus on the gamma-ray emission of hadronic origin.

\subsection{Leptonic emission}

As firstly pointed out by \citet{Kennel1984ApJ...283..710K}, the very different spectral indices in the radio and in the optical and above might indicate two populations of electrons generated by different physical processes, i.e. radio electrons and wind electrons (e.g. \citealt{Atoyan1996MNRAS.278..525A,Bandiera2002A&A...386.1044B,Meyer2010A&A...523A...2M}). While the latter are believed to be accelerated at the wind termination shock, the origin of the former has been linked to a relic population of the pulsar wind electrons \citep{Atoyan1999A&A...346L..49A}, or explained by acceleration in MHD turbulences by stochastic process and/or magnetic reconnection (e.g. \citealt{Nodes2004A&A...423...13N,Olmi2014MNRAS.438.1518O,Tanaka2017ApJ...841...78T,Lyutikov2019MNRAS.489.2403L,Luo2020ApJ...896..147L}). Here we do not discuss the origin of the radio component, but rather simply assume two electron populations according to the literature and the distribution of the radio one is given by
\begin{equation}
    n_{\text{r}}(r,\gamma_{\text{e}})=N_{\text{r},0}R_{\text{r}}^{-3}e^{-\frac{r^2}{2R_{\text{r}}^2}}\gamma_{\text{e}}^{-s_{\text{r}}} \quad \gamma_0\leq\gamma_{\text{e}}\leq\gamma_1,
\end{equation}
where the radial scale length $R_{\text{r}}=d_{\text{Crab}}\rho_{\text{r}}$ is independent of the Lorentz factor. We adopt the best-fitting values of $s_{\text{r}}=1.54$ and $\rho_{\text{r}}=89^{\prime\prime}$ from \citet{Dirson2023A&A...671A..67D}, and the similar energy range from $\gamma_0=20$ to $\gamma_1=9\times 10^4$. The normalization $N_{\text{r},0}$ is determined by the total energy of the radio electrons inside the nebula in order to fit the spectrum, which reaches $W_{\text{r}}=3.5\times 10^{48}$ erg and is consistent with \citet{Meyer2010A&A...523A...2M}. On the other hand, the distribution of the wind electrons is obtained by solving Eq. \ref{eq: distribution}.

\begin{figure*}[t]
    \centering
    \includegraphics[width=\linewidth]{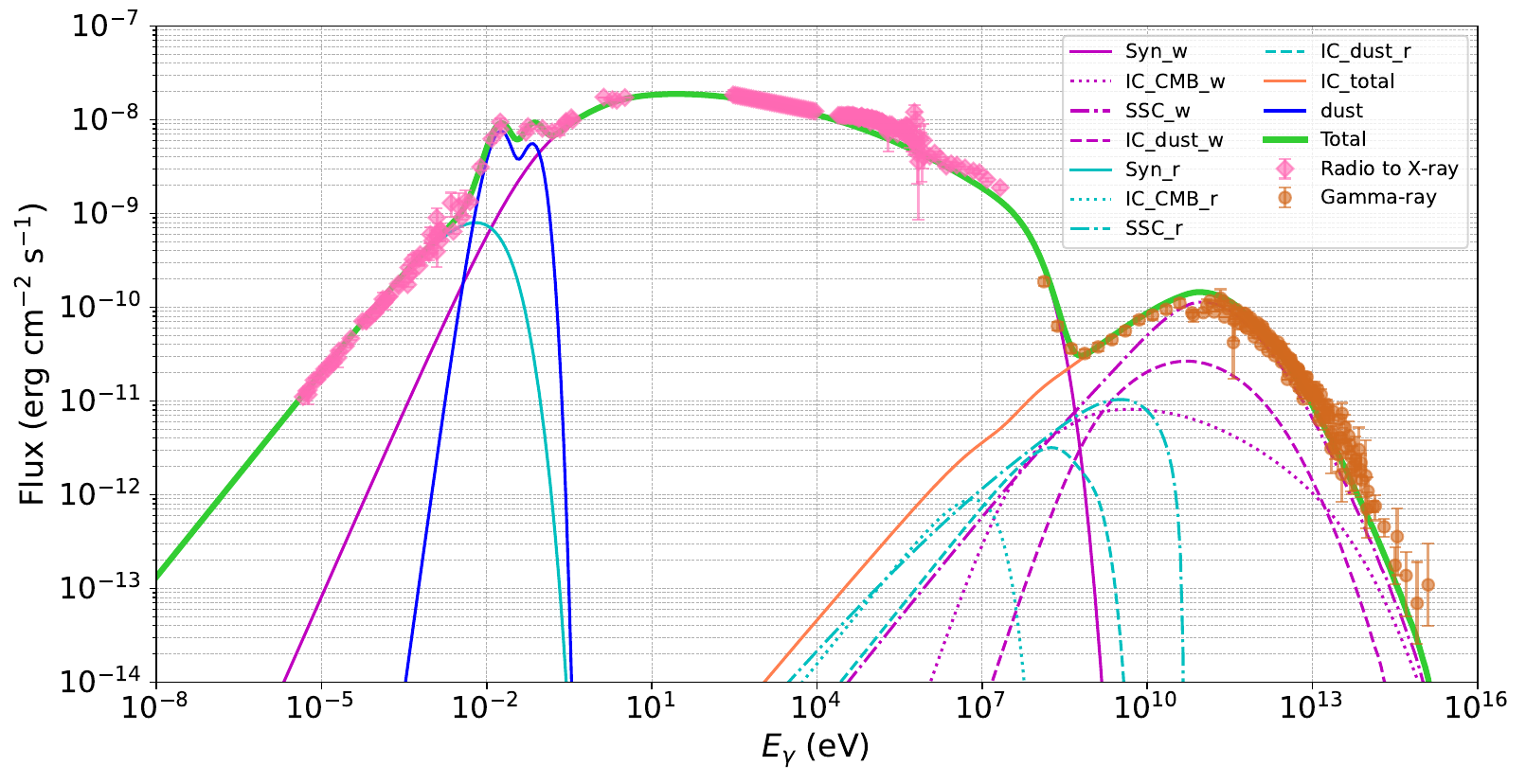}
    \caption{The fitting result of the multi-wavelength energy spectrum of the Crab nebula. The synchrotron (solid) and IC (CMB: dotted; synchrotron: dash-dotted; infrared: dashed) radiation from the radio (r; cyan) and wind (w; magenta) electrons is produced in the spatial-varying magnetic field and seed photon fields. The dust component (solid blue line) and the data from radio to X-ray (pink diamonds) are directly taken from \citet{Dirson2023A&A...671A..67D}. The gamma-ray data (brown circles) are collected from \textit{Fermi}-LAT \citep{Arakawa2020ApJ...897...33A}, HEGRA \citep{Aharonian2004ApJ...614..897A}, VERITAS \citep{Meagher2015ICRC...34..792M}, MAGIC \citep{Aleksic2015JHEAp...5...30A,Acciari2020A&A...635A.158M}, H.E.S.S. \citep{Aharonian2024A&A...686A.308A}, HAWC \citep{Abeysekara2019ApJ...881..134A}, Tibet AS+MD \citep{Amenomori2019PhRvL.123e1101A}, and LHAASO \citep{Cao2021Sci...373..425L}. Sum of the IC emission and of the emission across the whole spectrum are shown respectively with solid orange line and thick solid green line.
    }
    \label{fig: SED}
\end{figure*}

\begin{table}[t]
    \centering
    \caption{Values of Parameters for the Crab Nebula}
    \begin{tabular}{lcc}
        \hline
        \hline
        Parameter & Symbol & Value \\
        \hline
        \multicolumn{3}{c}{Fixed parameters} \\
        \hline
        SN explosion energy ($10^{51}$ erg) & $E_{\mathrm{sn}}$ & 1 \\
        Ejecta mass ($M_{\odot}$) & $M_{\mathrm{ej}}$ & 9 \\
        ISM density ($\mathrm{cm^{-3}}$) & $n_{\mathrm{ism}}$ & 0.1 \\
        Initial spin-down luminosity ($\mathrm{erg\,s^{-1}}$) & $L_0$ & $3\times10^{39}$ \\
        Initial spin-down timescale (yr) & $\tau_0$ & 680 \\
        Braking index & $n$ & 2.519 \\
        Age (yr) & $T_{\mathrm{age}}$ & 970 \\
        Distance (kpc) & $d_{\mathrm{Crab}}$ & 2 \\
        Profile index & $\beta$ & 0.5 \\
        \hline
        \multicolumn{3}{c}{Fitting parameters} \\
        \hline
        Low-energy power-law index & $\alpha_1$ & 1.7 \\
        High-energy power-law index & $\alpha_2$ & 2.3 \\
        Minimum Lorentz factor & $\gamma_{\mathrm{e}, \min}$ & $2\times 10^5$ \\
        Break Lorentz factor & $\gamma_{\mathrm{b}}$ & $1\times 10^6$ \\
        Magnetic fraction & $\eta_{\mathrm{B}}$ & 0.02 \\
        Electron fraction & $\eta_{\mathrm{e}}$ & 0.93 \\
        Proton fraction & $\eta_{\mathrm{p}}$ & 0.05 \\
        Magnetic field at TS ($\mathrm{\mu G}$) & $B_0$ & 234 \\
        Velocity factor & $V_f$ & 0.15 \\
        \hline
    \end{tabular}
    \label{tab: Crab}
\end{table}

We use the canonical $10^{51}$ erg for the SN explosion energy, and require the outer boundary $R_\text{pwn}$ to reach 2~pc at the present age, which corresponds to an ejecta mass of $9\,M_{\odot}$ in consistent with \citet{Bandiera2020MNRAS.499.2051B}. Note that the explosion energy is an open question, and some studies suggest a lower energy of $10^{50}$~erg (e.g. \citealt{Smith2013MNRAS.434..102S,Stockinger2020MNRAS.496.2039S,Temim2024ApJ...968L..18T}). Details of the parameters can be found in Table \ref{tab: Crab} and the result is shown in Figure \ref{fig: SED}. We obtain a magnetic field at the termination shock $B_0=234\,\mathrm{\mu G}$ with the profile index $\beta=0.5$ or correspondingly $\alpha=1-\beta=0.5$ in $B(r)=B_0(r/r_{\mathrm{ts}})^{-\alpha}$, similar to the best-fitting values of $B_0=264\pm9\,\mathrm{\mu G}$ with $\alpha=0.51\pm0.03$ from \citet{Dirson2023A&A...671A..67D} and $B_0=256\,\mathrm{\mu G}$ with $\alpha=0.47$ from \citet{Aharonian2024A&A...686A.308A}.

Another parameter of our concern is the proton fraction. As shown in Figure \ref{fig: SED}, data $\sim$~PeV from LHAASO suggest a possible hardening feature of the spectrum. This was explained with an additional population, which could be either leptonic or hadronic \citep{Cao2021Sci...373..425L}. The sub-dominant existence of protons or heavier nuclei has been proposed by several works before (e.g. \citealt{Atoyan1996MNRAS.278..525A,Bednarek1997PhRvL..79.2616B,Bednarek2003A&A...405..689B}), and gained increasing attention. \citet{Liu2021ApJ...922..221L} estimated the upper limit of $\eta_{\mathrm{p}}$ in the Crab nebula, suggesting that the maximum $\eta_{\mathrm{p}}$ allowed by the current LHAASO data may go up to $\sim$~(10~--~50)\%, considering the diffusive escape of particles. In this work, we assume that the PeV gamma-ray emission is mostly originated from the protons injected at the termination shock. To get the proton fraction, we can start with the magnetic fraction and the electron fraction which are constrained by the spectrum fitting, given that $\eta_{\mathrm{p}}=1-\eta_{\mathrm{B}}-\eta_{\mathrm{e}}$. Ratio between the energy density of the magnetic field and that of the radiation field is limited by observations. Once $R_{\mathrm{ts}}$ and $R_{\mathrm{pwn}}$ are given, the radiation fields are settled because of the fixed size and consequently the magnetic field. Therefore, there is little room to adjust $\eta_{\mathrm{B}}$ around 0.02. In principle, $\eta_{\mathrm{e}}$ could be as high as 0.98 if only electrons are considered \citep{Martin2022JHEAp..36..128M}, but it should not be lower than 0.9 as found in the phenomenological analyses \citep{Dirson2023A&A...671A..67D,Aharonian2024A&A...686A.308A}. After further tunning, we find that if $\eta_{\mathrm{e}}$ gets lower than 0.93, the synchrotron emission can hardly fit the optical and X-ray data. Hence, the maximum proton fraction allowed is 0.05, which is used in the following.

\subsection{Hadronic emission}

For the proton population, two values of injection index $\alpha_{\text{p}}$ are considered: the canonical one of 
2.0, which is suggested by the theory of diffusive shock acceleration (e.g. \citealt{Drury1983RPPh...46..973D}), and a harder one of 1.5. The minimum injection energy is assumed to be 1~TeV. Apart from the distribution obtained following procedures described in Section \ref{sec: method}, we still lack target gas density to derive the hadronic emission.

\begin{figure*}[t]
    \centering
    \includegraphics[width=\linewidth]{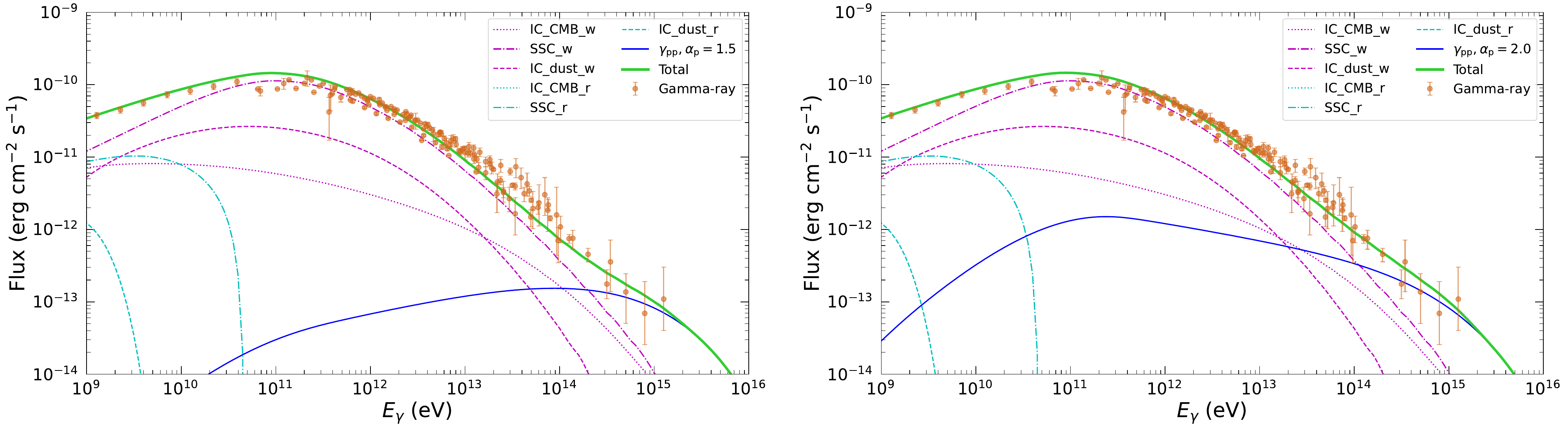}
    \caption{The spectral energy distribution above 1 GeV with proton population. Two injection indices $\alpha_{\text{p}}=1.5$ and 2.0 are considered, shown with solid blue lines in the left and right panels. The IC radiation from the electrons as well as gamma-ray data are the same as those in Figure \ref{fig: SED}.}
    \label{fig: amp}
\end{figure*}

The complex network of line-emitting filaments in the nebula is one of the most iconic features of the Crab. The interface between the PWN and the swept-up shell filled with ejecta is Rayleigh-Taylor (hereafter RT) unstable \citep{Chevalier1975ApJ...200..399C,Bandiera1983A&A...126....7B}. The "fingers" protruding into the PWN are expected to originate from the RT instability \citep{Hester1996ApJ...456..225H}, as shown by hydrodynamic and manetohydrodynamic simulations of the Crab \citep{Jun1998ApJ...499..282J,Bucciantini2004A&A...423..253B,Porth2014MNRAS.443..547P}. The density in the clumpy structures could be much higher than that of the ejecta \citep{Owen2015ApJ...801..141O}, making the straightforward assumption of a mean density an inaccurate choice. \citet{Atoyan1996MNRAS.278..525A} pointed out that particles that run into the over-density filaments would propagate slower, and the effective density would be significantly different from the mean gas density in the case of inhomogeneous distribution. Hence, it is a nontrivial task to determine the density used in the hadronic process.

As revealed by \citet{Porth2014MNRAS.443..547P}, the filaments growing from the PWN boundary do not fill the whole nebula but instead saturate at certain level. On the other hand, the relativistic wind continually blown by the central engine pushes the material outward. Therefore, the panorama of the gas distribution will be low density inside and high density outside. The extension of the filaments, or the saturation level, is about 40\%~$R_{\text{pwn}}$ shown by investigation of gas and dust distribution \citep{Owen2015ApJ...801..141O}, i.e. in the range (0.6~--~1)\,$R_{\text{pwn}}$. For simplicity, the nebula can be divided into two parts with different effective densities:
\begin{equation}
    n_{\text{eff}}(r)=
    \left\{
    \begin{array}{l}
        n_{\text{ej}} \quad r \leq 0.6R_{\mathrm{pwn}} \\
        f_a \times n_{\text{m}} \quad 0.6R_{\mathrm{pwn}} < r \leq R_{\mathrm{pwn}}
    \end{array}
    \right. , \label{eq: n_eff}
\end{equation}
where the ejecta density $n_{\text{ej}}=0.35\,\mathrm{cm}^{-3}$ obtained in Section \ref{sec: method} is adopted for the inner region. The total nebular mass is estimated to be $7.2\pm0.5\,M_{\odot}$ including a vast majority of $7.0\pm0.5\,M_{\odot}$ in gaseous form \citep{Owen2015ApJ...801..141O}, which is about 80\% of the total ejecta mass $9\,M_{\odot}$. The majority of the matter is loaded by the fall-back process via RT instability, and thus the simple assumption that the $0.8M_{\text{ej}}$ is distributed in the outer region is made in the following. The mean density in this region can be calculated as $n_{\text{m}}=0.8\,M_{\text{ej}}/V_{\text{outer}}/1.4m_{\text{p}}\approx8.3\,\mathrm{cm}^{-3}$. Note that the definition of amplification factor $f_a$ here is different from that in \citet{Atoyan1996MNRAS.278..525A}: in their work this quantity is defined as the ratio between the typical radius of the filaments and the characteristic scattering length of particles, which is energy-dependent, while here it is treated as a general effect of accumulation of the particles inside the filaments.

The only variable $f_a$ is adjusted by tuning the total flux at 1~PeV to reach $1\times10^{-13}$~erg~cm$^{-2}$~s$^{-1}$ so that the UHE gamma-ray data from LHAASO can be fitted with emission of hadronic origin. The fitting results with contribution from the protons are shown in Figure \ref{fig: amp}. In the two cases, the amplification factor $f_a\approx15$ for $\alpha_{\text{p}}=1.5$ and $f_a\approx60$ for $\alpha_{\text{p}}=2.0$, respectively.

\section{Neutrino emission from simulated population of young PWNe} \label{sec: population}

The Crab nebula has been treated as a standard template for the free-expanding PWN. In the following, we will use some results based on the spectrum fitting described in Section \ref{sec: Crab} to calculate the neutrino emission from the synthetic population of young PWNe in the Galaxy, grounded on the assumption that all the simulated sources evolve like the Crab. Details of the generating process will be introduced first, and then their neutrino emission is estimated and compared to the best-fit result from IceCube.

\subsection{Generating a pulsar population}

Previous works focusing on the population study have tried to simulate a pulsar sample with different treatments (e.g. \citealt{Watters2011ApJ...727..123W,Cristofari2017MNRAS.471..201C,Johnston2020MNRAS.497.1957J,Fiori2022MNRAS.511.1439F,Martin2022A&A...666A...7M,Chen2024MNRAS.527.7915C}). In general, information of four aspects should be taken into account: (i) evolution of the SNR, (ii) location of the pulsar, (iii) evolution of the PWN, and (iv) transport of the particles. According to \citet{Kasen2009ApJ...703.2205K}, the Type II SN explosion energy ranges from 0.5 to 4 $\times\,10^{51}$ erg spanning an order of magnitude. Here we consider a log-normal distribution of $E_\mathrm{sn}$ centering at the canonical value of $1\times10^{51}$ erg. The final stellar mass of the progenitor is related to its initial mass, metallicity, and rotation (e.g. \citealt{You2024ApJ...970..145Y}). Massive star less than 20 $M_{\odot}$ is believed to give birth to a neutron star after the SN explosion, while the more massive one may become a black hole \citep{Smartt2009ARA&A..47...63S}. For simplicity, we adopt a normal distribution which truncates at 5~$M_{\odot}$ \citep{Fiori2022MNRAS.511.1439F}, and values that over 15~$M_{\odot}$ are reset to 15~$M_{\odot}$, considering the requirement for core collapse and mass loss during stellar evolution. Midplane density of hydrogen gas in different radial regions from \citet{Lipari2018PhRvD..98d3003L} (see their Figure 5) is adopted for the ISM distribution. The SNR evolution is then determined with these three parameters.

\citet{Yusifov2004A&A...422..545Y} proposed a four-parameter Gamma function to depict the pulsar surface density, which has been updated with the latest observations \citep{Xie2024ApJ...963L..39X}:
\begin{equation}
    \rho(R) = A \left( \frac{R + R_{\mathrm{pdf}}}{R_{\odot} + R_{\mathrm{pdf}}} \right)^a \exp \left[ - b \left( \frac{R - R_{\odot}}{R_{\odot} + R_{\mathrm{pdf}}} \right) \right], \label{eq: surface density}
\end{equation}
where $R$ is the galactocentric radius, $R_{\odot} = 8.3$ kpc is the distance from the Sun to the Galactic center (GC), $A = 20.41\,\pm\,0.31$ kpc$^{-2}$, $a = 9.03\,\pm\,1.08$, $b = 13.99\,\pm\,1.36$, and $R_{\text{pdf}} = 3.76\,\pm\,0.42$. The surface density increases from the GC and reaches a maximum at a galactocentric radius of $\sim3.91$ kpc, and then gradually drops at larger distance. The probability density function for the radial distance $R$ can be obtained with the surface density given by Eq. \ref{eq: surface density}:
\begin{equation}
    P(R) \propto 2\pi R\rho(R). \label{eq: probability}
\end{equation}
Considering beaming correction proposed by \citet{Tauris1998MNRAS.298..625T}, the total number of pulsars in the Galaxy is about $(1.1\pm0.2)\times10^5$ \citep{Xie2024ApJ...963L..39X}.

It is expected that young pulsars of our concern mostly locate in spiral arms, as their OB star progenitors and the associated HII regions, giant molecular clouds and masers reveal the spiral structures (e.g. \citealt{Hou2014A&A...569A.125H,Chen2019MNRAS.487.1400C,Reid2019ApJ...885..131R}). There are five arms in total, including four major arms and the local arm, described by the following formula:
\begin{equation}
    \theta = \frac{\ln \left( \frac{R}{R_0} \right)}{\tan(\theta_1)} + \theta_0, \label{eq: theta}
\end{equation}
where $R$ and $\theta$ are polar coordinates centered at the GC, and $R_0$, $\theta_0$, and $\theta_1$ are the initial radii, the starting azimuth angle, and the pitch angle for the $i$th arm, respectively. The corresponding Cartesian coordinates are respectively $x=R\cos\theta$ and $y=R\sin\theta$ with axes parallel to $(l,b)=(90^{\circ},\,0^{\circ})$ and $(180^{\circ},\,0^{\circ})$. $\theta$ starts at the positive x-axis and increases counterclockwise, and the location of the Sun in the Cartesian frame is (0, 8.3 kpc). The parameters we use are listed in Table \ref{tab: spiral-arm}, which is based on \citet{Hou2014A&A...569A.125H} with a minor modification from \citet{Yao2017ApJ...835...29Y}. Note that the local arm is a sub-structure compared to other four gigantic arms, ending at $\theta\approx110^{\circ}$.

\begin{figure}[t]
    \centering
    \includegraphics[width=\linewidth]{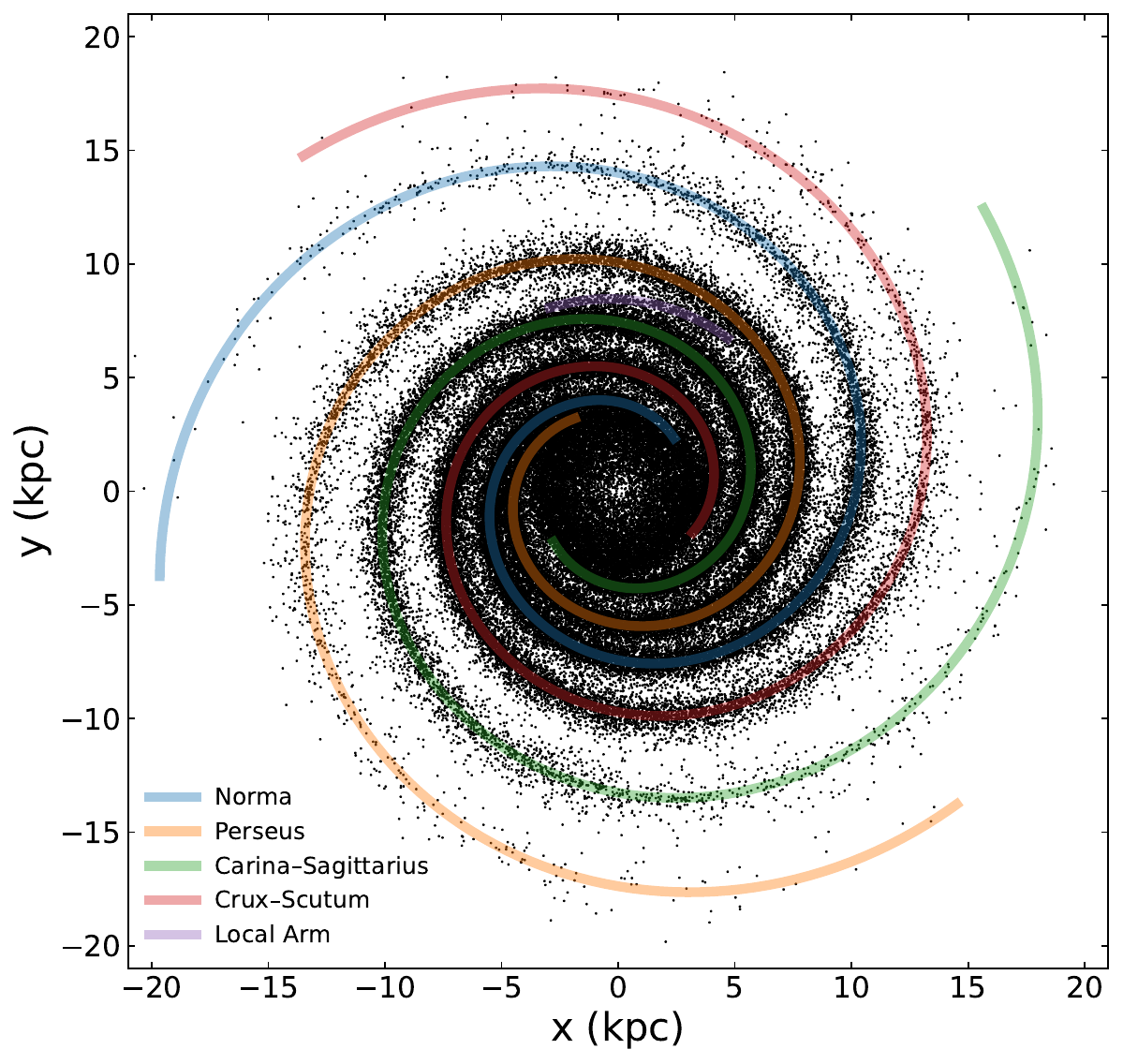}
    \caption{An example of the simulated distribution of pulsars in the $x-y$ plane. Each point represents the location of a pulsar on the Galactic plane. Solid lines with different colors trace the spiral arm centroids.}
    \label{fig: spiral-arm}
\end{figure}

\begin{table}[t]
    \centering
    \caption{Spiral-arm Parameters}
    \begin{tabular}{lcccc}
        \hline
        \hline
        Name & Index & $R_0$ (kpc) & $\theta_{0}$ (deg) & $\theta_1$ (deg) \\
        \hline
        Norma & 1  & 3.35 & 44.4 & 11.43 \\
        Perseus & 2 & 3.71 & 120.0 & 9.84 \\
        Carina--Sagittarius & 3 & 3.56 & 218.6 & 10.38 \\
        Crux--Scutum  & 4 & 3.67 & 330.3 & 10.54 \\
        Local Arm & 5  & 8.21 & 55.1 & 2.77 \\
        \hline
    \end{tabular}
    \label{tab: spiral-arm}
\end{table}

Following the procedure described in \citet{Xie2024ApJ...963L..39X}, the locations of the pulsars are obtained separately for the GC region and the spiral arms. In other words, a random index from Table \ref{tab: spiral-arm} or an additional "6" representing the GC is chosen for a pulsar. If it is located in the GC region within 3.71~kpc (the initial radius of the Perseus arm), the distance from the GC and the Galactic longitude of a simulated pulsar will be generated by two independent random processes, where the former follows Eq. \ref{eq: probability} and the latter is randomly chosen in [0, $2\pi$) rad. As for the spiral arms, the position depends on both the radial distribution of pulsars and the structure of the spiral arms. A synthesized pulsar is randomly distributed on the centroid of the $i$th spiral arm with a distance $R_{\text{raw}}$ chosen according to the radial distribution, and the corresponding polar angle $\theta_{\text{raw}}$ is calculated according to Eq. \ref{eq: theta}. To avoid artificial features, we use the method from \citet{Faucher-Giguere2006ApJ...643..332F} to blur the distribution: the polar angle of each pulsar in the region that the arms overlap with the GC, i.e. from 3.35 to 3.71 kpc, is corrected by applying $\theta_{\mathrm{corr}} \exp \left( -0.35 R_{\mathrm{raw}} / \mathrm{kpc} \right)$, where $\theta_{\mathrm{corr}}$ is randomly chosen in [0, $2\pi$) rad; pulsars in the spiral arms are further altered by adding a correction $R_{\mathrm{corr}}$ drawn from a normal distribution centered at zero with standard deviation 0.07$R_{\text{raw}}$, without preference with respect to direction. An example of the simulated birth distribution in the $x-y$ plane is illustrated in Figure \ref{fig: spiral-arm}. As we are only considering the young pulsars, we simply neglect their proper motions and thus all of them remain at their birth positions. Distance from the simulated pulsar to the Sun $d_{\mathrm{psr}}$ can then be obtained and subsequently used in the flux calculation.

The age of a pulsar $T_{\text{age}}$ is assigned to each source randomly with average birth rate of 1.5/century. The initial pulsar spin period $P_0$ is sampled from a normal distribution centered at 50~ms with a truncation at 10~ms, but the standard deviation is somehow arbitrary, including 10, 35 and 50~ms \citep{Watters2011ApJ...727..123W,Johnston2020MNRAS.497.1957J,Martin2022A&A...666A...7M}. Here we choose the middle one. The birth magnetic field at the surface of a pulsar $B_{\text{s}}$ is modeled with a log-normal distribution, in agreement with \citet{Faucher-Giguere2006ApJ...643..332F}. With these parameters and assuming pure dipole spin-down, namely $n = 3$, the initial spin-down luminosity as well as the initial spin-down timescale can be derived:
\begin{equation}
    L_0=\frac{B_{\mathrm{s}}^2R_{\mathrm{s}}^6}{6c^3}\left(\frac{2\pi}{P_0}\right)^4,
\end{equation}
\begin{equation}
    \tau_0 = \frac{3c^3IP_0^2}{4\pi^2B_{\mathrm{s}}^2R_{\mathrm{s}}^6},
\end{equation}
where the moment of inertia $I$ and the radius of a pulsar $R_{\mathrm{s}}$ have typical values of $10^{45}$~g~cm$^2$ and 12~km, respectively.

The last part is about the particles. Basically, we refer to the values obtained in the fitting process of the Crab nebula. $\eta_\text{B}$ and $\eta_\text{p}$ are fixed to the Crab values; the profile index $\beta$ is again fixed to 0.5; the velocity factor $V_f$ adopts the mean value between 0 and 1. We summarize all of the input parameters mentioned above in Table \ref{tab: sam}.

\begin{table*}[t]
    \centering
    \caption{Summary of the input parameters used to generate the young PWNe population.}
    \begin{tabular}{lll}
        \hline
        \hline
        Parameter & Distribution & Value \\
        \hline
        \multicolumn{3}{l}{\textbf{Parameters for the SNRs}} \\
        SN explosion energy $E_{\mathrm{sn}}$ (erg) & Log$_{10}$-normal & $\mu=51$, $\sigma=0.2$ \\
        Ejecta mass $M_{\mathrm{ej}}$ ($M_{\odot}$) & Normal & $\mu=10$, $\sigma=2$, truncated at 5, values $\geq\,15$ reset to 15 \\
        ISM density $n_{\mathrm{ism}}$ (cm$^{-3}$) & \multicolumn{2}{l}{Following \citet{Lipari2018PhRvD..98d3003L}} \\
        \hline
        \multicolumn{3}{l}{\textbf{Parameters for the PSRs}} \\
        Braking index $n$ & Constant & $3$ \\
        Surface magnetic field at birth $B_{\mathrm{s}}$ (G) & Log$_{10}$-normal & $\mu = 12.65$, $\sigma = 0.55$ \\
        Initial spin period $P_0$ (ms) & normal & $\mu = 50$, $\sigma = 35$, truncated at 10 \\
        \hline
        \multicolumn{3}{l}{\textbf{Parameters for the particles}} \\
        Magnetic fraction $\eta_\mathrm{B}$ & Constant & $0.02$ \\
        Proton fraction $\eta_\mathrm{p}$ & Constant & $0.05$ \\
        Profile index $\beta$ & Constant & $0.5$ \\
        Velocity factor $V_f$ & Constant & $0.5$ \\
        \hline
    \end{tabular}
    \label{tab: sam}
\end{table*}

\subsection{Predicted neutrino emission}

Before taking the next step, it should be noted that not all simulated PWNe we generate will contribute to the flux measured on Earth. Some of them may be too distant to be detected; some of them can be seen but have already entered the later phase of evolution. If we require $R_{\mathrm{pwn}}=R_{\mathrm{rs}}$, which are given by Eq. \ref{eq: R_pwn} and \ref{eq: R_rs}, then the period of the free expansion phase $T_{\text{free}}$ is determined. The propagation time of the emission is easily obtained with $T_{\text{prop}}=d_{\text{psr}}/c$. Therefore, sources will be taken into consideration only if $T_{\text{age}}-T_{\text{prop}}=T_{\text{obs}}>0$ and $T_{\text{obs}}\leq T_{\text{free}}$, as the former ensures that the emission can arrive and the latter guarantees a young PWN of our concern.

The ones that pass the selection criteria above then follow the calculation procedures described in Section \ref{sec: method}. For the target gas density, we make the following assumptions: sources with $T_{\text{obs}}<500$~years has only one density value for the whole area, i.e. $n_\text{eff}=n_{\text{ej}}$, while the older counterparts have the double density structure similar to the Crab (see Eq. \ref{eq: n_eff}). For the latter case, sources with $T_{\text{obs}}>1000$~years use the parameters obtained from the Crab fitting, i.e. the saturation level of the filaments is set to be 0.4, corresponding to 80\% of the ejecta; for the younger ones aging between (500, 1000]~years, the saturation level takes a lower value of 0.2 \citep{Porth2014MNRAS.443..547P}, and the mass fraction of fall-back ejecta is assumed to be 40\%; the amplification factor uses $f_a\approx15$ for $\alpha_{\text{p}}=1.5$ and $f_a\approx60$ for $\alpha_{\text{p}}=2.0$. The aim of the age division is to imitate the growth of the filaments over time in a simplified way. 

The last thing to mention is the removal of some sources with extreme parameters. The first criterion is that young sources that are less than 50~years are excluded. There are mainly two reasons: the applicability of the model is uncertain at the very beginning of the system; observationally no recent SN has been found in our Galaxy. Another condition is about the observed flux. If all the spin-down power is fully converted into emission, then the flux from the Crab without attenuation at present will be $F=L/4\pi d^2\sim 10^{-6}$~erg~cm$^{-2}$~s$^{-1}$ using Eq. \ref{eq: spin-down L} and parameters in Table \ref{tab: Crab}. The simulated sources with $F>10^{-5}$~erg~cm$^{-2}$~s$^{-1}$ are removed optionally, while the age criterion is compulsory in the calculation. We will discuss these \textit{a priori} arguments later.

\begin{figure*}
    \centering
    \includegraphics[width=\linewidth]{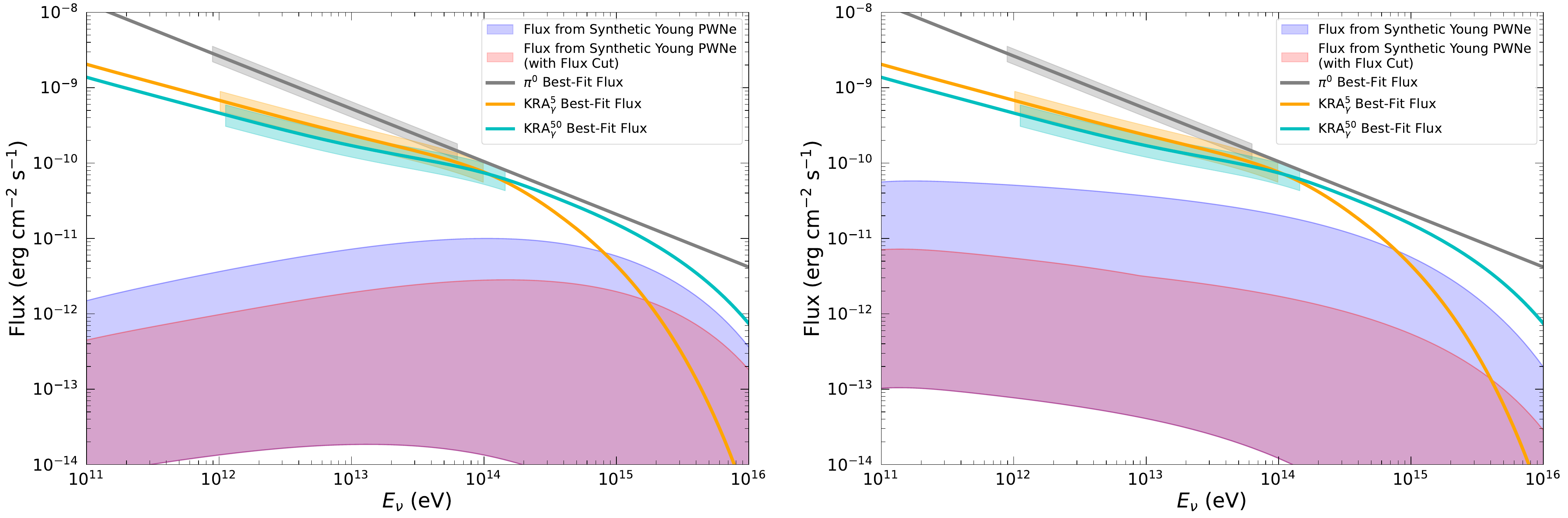}
    \caption{Predicted all-flavor neutrino flux from the synthetic young PWNe population. The left and right panels are the cases with proton injection indices being 1.5 and 2.0, respectively. The combination of the blue and purple bands is the neutrino flux from the simulated sources, while the purple region alone is the result with additional flux criterion. The upper and lower bounds are correspondingly the maximum and minimum of the whole population. The shaded bands in gray, orange and cyan are respectively the best-fit results with $1\sigma$ uncertainties of $\pi^0$, KRA$_{\gamma}^{5}$ and KRA$_{\gamma}^{50}$ models from \citet{Abbasi2023Sci...380.1338I}.}
    \label{fig: v flux}
\end{figure*}

For each injection index, we repeat the whole produces for 100 times and the total number of the retained sources is 4918, which corresponds to about 50 pulsars per run. Among the sample 71 sources are removed according to the 50-year criterion. The all-flavor neutrino flux from the simulated population after age cut is shown in Figure \ref{fig: v flux} together with the best-fit results from \citet{Abbasi2023Sci...380.1338I}. The upper and lower bounds are the maximum and minimum values of the predicted flux, and the blue and purple bands representing results before and after flux cut share the lower bound.

The predicted neutrino flux is $\sim1\times10^{-11}$~erg~cm$^{-2}$~s$^{-1}$ for $\alpha_{\text{p}}=1.5$ and $\sim2\times10^{-11}$~erg~cm$^{-2}$~s$^{-1}$ for $\alpha_{\text{p}}=2.0$ at 100~TeV. When the flux criterion is further adopted, only 3 sources will be excluded but the upper bound will drop to $\sim3\times10^{-12}$~erg~cm$^{-2}$~s$^{-1}$ and $\sim2\times10^{-12}$~erg~cm$^{-2}$~s$^{-1}$, which is $\sim5\%$ relative to the lower limit of the KRA$_{\gamma}^{50}$ model. The prominent contribution from these 3 sources originates from the combination of high luminosity and close distance. In the higher PeV energy range, the three diffuse models differentiate from each other. Due to the cutoff energy at 5~PeV, the KRA$_{\gamma}^{5}$ model declines rapidly and even becomes a bit lower than the predicted flux from the young PWNe at several PeV. Meanwhile, the KRA$_{\gamma}^{50}$ model and the simple extrapolation of the $\pi^0$ model approach each other around $2\times10^{-11}$~erg~cm$^{-2}$~s$^{-1}$ at 1~PeV. The two cases with different injection indices both mildly drop to $\sim6\times10^{-12}$~erg~cm$^{-2}$~s$^{-1}$. Considering the flux cut, emission from $\alpha_{\text{p}}=1.5$ hardly changes compared to value at 100~TeV, and it is nearly quadruple the descending $\alpha_{\text{p}}=2.0$. Being $\sim45\%$ of the KRA$_{\gamma}^{5}$ model, and $\sim10\%$ of the KRA$_{\gamma}^{50}$ model and the $\pi^0$ model, the case of $\alpha_{\text{p}}=1.5$ indicates increasing contribution from discrete sources.

\section{Discussion} \label{sec: dis}

\subsection{Comparison with previous works of Crab fitting}

In Section \ref{sec: Crab}, we fit the multi-wavelength spectrum of the Crab nebula. We note that \citet{Peng2022ApJ...926....7P} used a similar dynamic model to fit the spectrum, but the two results obtain different values for the same parameter. Here are the major discrepancies in the settings.
\begin{itemize}
    \item[(i)] While synchrotron and IC emission in our work originates from both radio and wind electrons, following \citet{Dirson2023A&A...671A..67D} and \citet{Aharonian2024A&A...686A.308A}, only electrons injected from the termination shock are considered in \citet{Peng2022ApJ...926....7P}.
    \item[(ii)] We adopt the dust component from \citet{Dirson2023A&A...671A..67D}, and the associated IR radiation field is inhomogeneous at different radii in the PWN according to Eq.~(\ref{eq: photon seed}). \citet{Peng2022ApJ...926....7P} instead considered homogeneous fields in NIR and FIR.
    \item[(iii)] We calculate the synchrotron self-Compton (SSC) radiation with seed photons again using Eq.~(\ref{eq: photon seed}). Assuming homogeneous distribution of this radiation in the spherical volume, \citet{Peng2022ApJ...926....7P} used an approximation\footnote{Note, however, equations (the two without indices beyond Eq. 24) given in \citet{Peng2022ApJ...926....7P} are incorrect as they use $R_{\text{ts}}$ instead of $R_{\text{pwn}}$.} (see Section 4.1 in \citet{Atoyan1996MNRAS.278..525A}): 
    \begin{equation}
        n_{\text{syn}}(r, \epsilon) = \frac{Q_{\text{syn}}(\epsilon)}{4 \pi c R_{\mathrm{pwn}}^2} U(x),
    \end{equation}
    where $x=r/R_{\mathrm{pwn}}$, and
    \begin{equation}
        U(x) = \frac{3}{2} \int_0^1 \mathrm{d}y\,\frac{y}{x} \ln \frac{x+y}{|x-y|}.
    \end{equation}
\end{itemize}
These factors lead to two main differences. Firstly, we can simply compare the timescales of advection and diffusion respectively defined as $\tau_{\text{adv}}\sim R_{\text{pwn}}\,/\,V_{\text{ts}}$ and $\tau_{\text{diff}}\sim R_{\text{pwn}}^2\,/\,D_{\text{ts, 100 TeV}}$, where the latter is calculated for particles with $E=100$~TeV. In \citet{Peng2022ApJ...926....7P}, $\tau_{\text{diff}}$ is about two orders of magnitude smaller than $\tau_{\text{adv}}$, which indicates dominating diffuse propagation. In our case, however, $\tau_{\text{diff}}\sim\tau_{\text{adv}}$, implying competitive relation between the two channels around 100~TeV. Only in sub-PeV to PeV range does diffusion start to take over. Our case is closer to the \textit{IXPE} result \citep{Bucciantini2023NatAs...7..602B}, where particles are subject to the toroidal magnetic field. Secondly, \citet{Peng2022ApJ...926....7P} obtained $\eta_{\mathrm{B}}=0.06$, $\eta_{\mathrm{e}}=0.7$, and $\eta_{\mathrm{p}}=0.24$. Regardless of the existence of protons, our result is consistent with the dominant electron fraction $\eta_{\mathrm{e}}\gtrsim0.9$ found in recent works of Crab spectrum fitting (e.g. \citealt{Martin2022JHEAp..36..128M,Dirson2023A&A...671A..67D,Aharonian2024A&A...686A.308A}).

\citet{Martin2022JHEAp..36..128M} solved a different time-dependent transport equation to obtain the particle distribution. Like \citet{Peng2022ApJ...926....7P}, radio electrons and inhomogeneous IR radiation field were not included. The Crab spectrum was fitted without protons (i.e. $\eta_{\text{B}}+\eta_{\text{e}}=1$), and therefore there was no hardening feature around PeV. Though they did not consider the radial dependence of the magnetic field, $\eta_{\text{B}}=0.02$ in their work is consistent with our result.

\subsection{Reflections on predicted neutrino emission}

In the synthetic population, part of the sources are removed according to the age criterion and flux criterion. If the 71 sources with age less than 50~years are not excluded, the flux from the simulated pulsars will overshoot the best-fit results from IceCube, confirming the necessity of the age cut. As for the flux constraint, the removed 3 sources make significant contribution to the predicted neutrino flux. Even though the criterion itself ($F>10F_{\text{Crab}}$) seems somehow arbitrary, it heuristically suggests that the flux we measure on Earth may be attributed to several strong discrete sources, especially in the PeV regime. Distribution of PeV cosmic rays in our Galaxy is found to be significantly clumpy and inhomogeneous and different from the GeV counterparts \citep{Giacinti2023arXiv230510251G}, while the neutrino emission from TeV to PeV may also be associated to particular regions in the Galaxy, e.g. the local bubble \citep{Bouyahiaoui2020PhRvD.101l3023B}, the Galactic ridge \citep{Albert2023PhLB..84137951A,Neronov2023PhRvD.108j3044N}, and the Cygnus X region \citep{LHAASO2024SciBu..69..449L}.

In reality, the structure of the filaments can be rather complex, and the inhomogeneity shown by simulations is rooted in the SN explosion \citep{Jun1998ApJ...499..282J} and/or the injection from the pulsar \citep{Porth2014MNRAS.443..547P}. Our treatment of the filaments in the nebula is rather simplified, as the amplification factor $f_a$ only represents a general effect of accumulation and does not vary over time for a given injection index. The sample is separated into three categories according to the observational age of the source, but the actual growth of the structure is nontrivial due to the existence of linear and non-linear stages of the RT instability (see e.g. \citealt{Porth2014MNRAS.443..547P}). More observations of the filaments as well as precise gamma-ray measurement will definitely help constrain this factor, but the number of ideally observable young PWNe is limited as shown by our simulated population. Hence, dedicated simulations of filaments will be important to shed light on the amplification effect.

We mainly focus on the free-expanding PWN because of the confinement of the toroidal magnetic field and the massive injection from the young pulsar. When the reverse shock interacts with the PWN, the toroidal configuration may be disrupted, leading to efficient escape due to diffusion \citep{Hinton2011ApJ...743L...7H}. On the other hand, the compression of PWN may result in a denser environment and enhance the production of neutrinos. The number of PWNe in the reverberation phase may probably be more abundant than that in the free-expanding phase, as the former phase may last a longer duration than the latter one in particular for those energetic pulsars \citep{Bandiera2023MNRAS.520.2451B,Bandiera2023MNRAS.525.2839B}. Therefore, potential high-energy neutrino production in this stage deserves further investigations.

In the sub-PeV to PeV range, the neutrino emission originated from the cosmic-ray sea strongly depends on the diffuse template \citep{Abbasi2023Sci...380.1338I}. The simple extrapolation of the $\pi^0$ model maintain at a high level, while the prediction of KRA$_{\gamma}$ model is affected by the cutoff energy \citep{Gaggero2015ApJ...815L..25G}. Recently, Baikal-GVD also found an excess of neutrinos with 8 cascade events from low Galactic latitudes above 200 TeV \citep{Allakhverdyan2024arXiv241105608A}. However, their result in the range from  200~TeV to 1~PeV is higher than the extrapolation of IceCube, challenging contemporary scenarios of cosmic-ray templates (see also discussion in \citealt{Troitsky2024PhyU...67..349T}). On the other hand, other types of discrete sources besides young PWNe could be possible contributors as well, according to catalog stacking analyses (e.g. \citealt{Gagliardini2024ApJ...969..161G}). With the development of KM3NeT and Baikal-GVD as well as the construction of new detectors like IceCube-Gen2 \citep{Aartsen2021JPhG...48f0501A}, P-ONE \citep{Agostini2020NatAs...4..913A}, TRIDENT \citep{Ye2023NatAs...7.1497Y} and HUNT \citep{Huang2024icrc.confE1080H}, the mystery of Galactic neutrino emission may be unveiled in the near future.

\section{Conclusion} \label{sec: con}

The Milky Way has been shown to be a neutrino source in the sky. Previous researches suggest dominant diffuse emission in the TeV-PeV range, but contribution from sources in the Galaxy cannot be omitted. PWNe, especially the young ones that are still in the free expansion phase, have been regarded as possible contributors with estimation of neutrino emission from individual sources as well as stacking analyses using different catalogs. In this work, instead of following these conventional methods, we  directly calculate the neutrino emission from a synthetic young PWNe population.

A dynamic model is employed to depict the evolution of the free-expansion PWN, and the distribution of the relativistic particles is obtained by solving a transport equation with temporal and spatial evolution. The Crab nebula is treated as a standard template, whose multi-wavelength spectrum is overall fitted by synchrotron and IC radiation from radio and wind electrons. The magnetic field follows a power-law distribution $B(r)=B_0\,(r/R_{\text{ts}})^{-0.5}$, where $B_0=234\,\mathrm{\mu G}$ is the value at present. UHE gamma-ray emission around 1~PeV is mainly expected to originate from the hadronic population, satisfying $\eta_{\text{B}}+\eta_{\text{e}}+\eta_{\text{p}}=1$. The nebula is simply divided into an inner spherical region of low density and an outer one filled with filaments. 80\% of the ejecta is assumed to fall back into the outer area (0.6~--~1)\,$R_{\text{pwn}}$ due to the RT instability. By tuning the total flux at 1~PeV to reach $1\times10^{-13}$~erg~cm$^{-2}$~s$^{-1}$ revealed by LHAASO, the amplification factor of the filaments is determined for two injection indices: $f_a\approx15$ for $\alpha_{\text{p}}=1.5$ and $f_a\approx60$ for $\alpha_{\text{p}}=2.0$.

To estimate the neutrino emission from the young PWNe in the Galaxy, a synthetic population is generated on the assumption that all sources evolve with the same energy partition as the Crab. The simulated sources are assumed to locate in the Galactic plane with spiral structure taken into account. After excluding sources whose observed ages are less than 50~years and sources that emit extremely strong flux an order of magnitude higher than the Crab, the neutrino emission from the simulated young PWNe is found to be about 5\% of the best-fit results from IceCube for both injection indices at 100~TeV in the optimistic case. At the higher 1~PeV, emission from the scenario of $\alpha_{\text{p}}=2.0$ drops quickly while the harder $\alpha_{\text{p}}=1.5$ change mildly. On the other hand, total emission strongly depends on the template. Flux from the KRA$_{\gamma}^{5}$ model becomes lower than that from the synthetic population at several PeV due to the early cutoff, while $\pi^0$ and KRA$_{\gamma}^{50}$ models remain beyond the source contribution. More data collected from worldwide detectors and more precise cosmic-ray template are needed to investigate the origin of Galactic neutrino emission.

\section*{Acknowledgments}

We thank Kai Yan for the help with the KRA$\gamma$ models. This work is
supported by the National Natural Science Foundation of China under grants Nos. 12393852, 12333006 and 12121003.

\bibliography{sample631}{}

\begin{thebibliography}{}
\expandafter\ifx\csname natexlab\endcsname\relax\def\natexlab#1{#1}\fi
\providecommand{\url}[1]{\href{#1}{#1}}
\providecommand{\dodoi}[1]{doi:~\href{http://doi.org/#1}{\nolinkurl{#1}}}
\providecommand{\doeprint}[1]{\href{http://ascl.net/#1}{\nolinkurl{http://ascl.net/#1}}}
\providecommand{\doarXiv}[1]{\href{https://arxiv.org/abs/#1}{\nolinkurl{https://arxiv.org/abs/#1}}}

\bibitem[{{Aartsen} {et~al.}(2017{\natexlab{a}}){Aartsen}, {Ackermann}, {Adams}, {Aguilar}, {Ahlers}, {Ahrens}, {Altmann}, {Andeen}, {Anderson}, {Ansseau}, {Anton}, {Archinger}, {Arg{\"u}elles}, {Auer}, {Auffenberg}, {Axani}, {Baccus}, {Bai}, {Barnet}, {Barwick}, {Baum}, {Bay}, {Beattie}, {Beatty}, {Becker Tjus}, {Becker}, {Bendfelt}, {BenZvi}, {Berley}, {Bernardini}, {Bernhard}, {Besson}, {Binder}, {Bindig}, {Bissok}, {Blaufuss}, {Blot}, {Boersma}, {Bohm}, {B{\"o}rner}, {Bos}, {Bose}, {B{\"o}ser}, {Botner}, {Bouchta}, {Braun}, {Brayeur}, {Bretz}, {Bron}, {Burgman}, {Burreson}, {Carver}, {Casier}, {Cheung}, {Chirkin}, {Christov}, {Clark}, {Classen}, {Coenders}, {Collin}, {Conrad}, {Cowen}, {Cross}, {Day}, {Day}, {de Andr{\'e}}, {De Clercq}, {del Pino Rosendo}, {Dembinski}, {De Ridder}, {Descamps}, {Desiati}, {de Vries}, {de Wasseige}, {de With}, {DeYoung}, {D{\'\i}az-V{\'e}lez}, {di Lorenzo}, {Dujmovic}, {Dumm}, {Dunkman}, {Eberhardt}, {Edwards}, {Ehrhardt}, {Eichmann}, {Eller}, {Euler}, {Evenson}, {Fahey}, {Fazely}, {Feintzeig}, {Felde}, {Filimonov}, {Finley}, {Flis}, {F{\"o}sig}, {Franckowiak}, {Fr{\`e}re}, {Friedman}, {Fuchs}, {Gaisser}, {Gallagher}, {Gerhardt}, {Ghorbani}, {Giang}, {Gladstone}, {Glauch}, {Glowacki}, {Gl{\"u}senkamp}, {Goldschmidt}, {Gonzalez}, {Grant}, {Griffith}, {Gustafsson}, {Haack}, {Hallgren}, {Halzen}, {Hansen}, {Hansmann}, {Hanson}, {Haugen}, {Hebecker}, {Heereman}, {Helbing}, {Hellauer}, {Heller}, {Hickford}, {Hignight}, {Hill}, {Hoffman}, {Hoffmann}, {Hoshina}, {Huang}, {Huber}, {Hulth}, {Hultqvist}, {In}, {Inaba}, {Ishihara}, {Jacobi}, {Jacobsen}, {Japaridze}, {Jeong}, {Jero}, {Jones}, {Jones}, {Joseph}, {Kang}, {Kappes}, {Karg}, {Karle}, {Katz}, {Kauer}, {Keivani}, {Kelley}, {Kemp}, {Kheirandish}, {Kim}, {Kim}, {Kintscher}, {Kiryluk}, {Kitamura}, {Kittler}, {Klein}, {Kleinfelder}, {Kleist}, {Kohnen}, {Koirala}, {Kolanoski}, {Konietz}, {K{\"o}pke}, {Kopper}, {Kopper}, {Koskinen}, {Kowalski}, {Krasberg}, {Krings}, {Kroll}, {Kr{\"u}ckl}, {Kr{\"u}ger}, {Kunnen}, {Kunwar}, {Kurahashi}, {Kuwabara}, {Labare}, {Laihem}, {Landsman}, {Lanfranchi}, {Larson}, {Lauber}, {Laundrie}, {Lennarz}, {Leich}, {Lesiak-Bzdak}, {Leuermann}, {Lu}, {Ludwig}, {L{\"u}nemann}, {Mackenzie}, {Madsen}, {Maggi}, {Mahn}, {Mancina}, {Mandelartz}, {Maruyama}, {Mase}, {Matis}, {Maunu}, {McNally}, {McParland}, {Meade}, {Meagher}, {Medici}, {Meier}, {Meli}, {Menne}, {Merino}, {Meures}, {Miarecki}, {Minor}, {Montaruli}, {Moulai}, {Murray}, {Nahnhauer}, {Naumann}, {Neer}, {Newcomb}, {Niederhausen}, {Nowicki}, {Nygren}, {Obertacke Pollmann}, {Olivas}, {O'Murchadha}, {Palczewski}, {Pandya}, {Pankova}, {Patton}, {Peiffer}, {Penek}, {Pepper}, {P{\'e}rez de los Heros}, {Pettersen}, {Pieloth}, {Pinat}, {Price}, {Przybylski}, {Quinnan}, {Raab}, {R{\"a}del}, {Rameez}, {Rawlins}, {Reimann}, {Relethford}, {Relich}, {Resconi}, {Rhode}, {Richman}, {Riedel}, {Robertson}, {Rongen}, {Roucelle}, {Rott}, {Ruhe}, {Ryckbosch}, {Rysewyk}, {Sabbatini}, {Sanchez Herrera}, {Sandrock}, {Sandroos}, {Sandstrom}, {Sarkar}, {Satalecka}, {Schlunder}, {Schmidt}, {Schoenen}, {Sch{\"o}neberg}, {Schukraft}, {Schumacher}, {Seckel}, {Seunarine}, {Solarz}, {Soldin}, {Song}, {Spiczak}, {Spiering}, {Stanev}, {Stasik}, {Stettner}, {Steuer}, {Stezelberger}, {Stokstad}, {St{\"o}{\ss}l}, {Str{\"o}m}, {Strotjohann}, {Sulanke}, {Sullivan}, {Sutherland}, {Taavola}, {Taboada}, {Tatar}, {Tenholt}, {Ter-Antonyan}, {Terliuk}, {Te{\v{s}}i{\'c}}, {Thollander}, {Tilav}, {Toale}, {Tobin}, {Toscano}, {Tosi}, {Tselengidou}, {Turcati}, {Unger}, {Usner}, {Vandenbroucke}, {van Eijndhoven}, {Vanheule}, {van Rossem}, {van Santen}, {Vehring}, {Voge}, {Vogel}, {Vraeghe}, {Wahl}, {Walck}, {Wallace}, {Wallraff}, {Wandkowsky}, {Weaver}, {Weiss}, {Wendt}, {Westerhoff}, {Wharton}, {Whelan}, {Wickmann}, {Wiebe}, {Wiebusch}, {Wille}, {Williams}, {Wills}, {Wisniewski}, {Wolf}, {Wood}, {Woolsey}, {Woschnagg}, {Xu}, {Xu}, {Xu}, {Yanez}, {Yodh}, {Yoshida}, \& {Zoll}}]{Aartsen2017JInst..12P3012A}
{Aartsen}, M.~G., {Ackermann}, M., {Adams}, J., {et~al.} 2017{\natexlab{a}}, Journal of Instrumentation, 12, P03012, \dodoi{10.1088/1748-0221/12/03/P03012}

\bibitem[{{Aartsen} {et~al.}(2017{\natexlab{b}}){Aartsen}, {Ackermann}, {Adams}, {Aguilar}, {Ahlers}, {Ahrens}, {Samarai}, {Altmann}, {Andeen}, {Anderson}, {Ansseau}, {Anton}, {Arg{\"u}elles}, {Auffenberg}, {Axani}, {Bagherpour}, {Bai}, {Barron}, {Barwick}, {Baum}, {Bay}, {Beatty}, {Becker Tjus}, {Becker}, {BenZvi}, {Berley}, {Bernardini}, {Besson}, {Binder}, {Bindig}, {Blaufuss}, {Blot}, {Bohm}, {B{\"o}rner}, {Bos}, {Bose}, {B{\"o}ser}, {Botner}, {Bourbeau}, {Bradascio}, {Braun}, {Brayeur}, {Brenzke}, {Bretz}, {Bron}, {Burgman}, {Carver}, {Casey}, {Casier}, {Cheung}, {Chirkin}, {Christov}, {Clark}, {Classen}, {Coenders}, {Collin}, {Conrad}, {Cowen}, {Cross}, {Day}, {de Andr{\'e}}, {De Clercq}, {DeLaunay}, {Dembinski}, {De Ridder}, {Desiati}, {de Vries}, {de Wasseige}, {de With}, {DeYoung}, {D{\'\i}az-V{\'e}lez}, {di Lorenzo}, {Dujmovic}, {Dumm}, {Dunkman}, {Eberhardt}, {Ehrhardt}, {Eichmann}, {Eller}, {Evenson}, {Fahey}, {Fazely}, {Felde}, {Filimonov}, {Finley}, {Flis}, {Franckowiak}, {Friedman}, {Fuchs}, {Gaisser}, {Gallagher}, {Gerhardt}, {Ghorbani}, {Giang}, {Glauch}, {Gl{\"u}senkamp}, {Goldschmidt}, {Gonzalez}, {Grant}, {Griffith}, {Haack}, {Hallgren}, {Halzen}, {Hanson}, {Hebecker}, {Heereman}, {Helbing}, {Hellauer}, {Hickford}, {Hignight}, {Hill}, {Hoffman}, {Hoffmann}, {Hokanson-Fasig}, {Hoshina}, {Huang}, {Huber}, {Hultqvist}, {In}, {Ishihara}, {Jacobi}, {Japaridze}, {Jeong}, {Jero}, {Jones}, {Kalacynski}, {Kang}, {Kappes}, {Karg}, {Karle}, {Katz}, {Kauer}, {Keivani}, {Kelley}, {Kheirandish}, {Kim}, {Kim}, {Kintscher}, {Kiryluk}, {Kittler}, {Klein}, {Kohnen}, {Koirala}, {Kolanoski}, {K{\"o}pke}, {Kopper}, {Kopper}, {Koschinsky}, {Koskinen}, {Kowalski}, {Krings}, {Kroll}, {Kr{\"u}ckl}, {Kunnen}, {Kunwar}, {Kurahashi}, {Kuwabara}, {Kyriacou}, {Labare}, {Lanfranchi}, {Larson}, {Lauber}, {Lennarz}, {Lesiak-Bzdak}, {Leuermann}, {Liu}, {Lu}, {L{\"u}nemann}, {Luszczak}, {Madsen}, {Maggi}, {Mahn}, {Mancina}, {Maruyama}, {Mase}, {Maunu}, {McNally}, {Meagher}, {Medici}, {Meier}, {Menne}, {Merino}, {Meures}, {Miarecki}, {Micallef}, {Moment{\'e}}, {Montaruli}, {Moore}, {Moulai}, {Nahnhauer}, {Nakarmi}, {Naumann}, {Neer}, {Niederhausen}, {Nowicki}, {Nygren}, {Obertacke Pollmann}, {Olivas}, {O'Murchadha}, \& {Palczewski}}]{Aartsen2017ApJ...849...67A}
---. 2017{\natexlab{b}}, \apj, 849, 67, \dodoi{10.3847/1538-4357/aa8dfb}

\bibitem[{{Aartsen} {et~al.}(2020){Aartsen}, {Ackermann}, {Adams}, {Aguilar}, {Ahlers}, {Ahrens}, {Alispach}, {Andeen}, {Anderson}, {Ansseau}, {Anton}, {Arg{\"u}elles}, {Auffenberg}, {Axani}, {Bagherpour}, {Bai}, {Balagopal V.}, {Barbano}, {Barwick}, {Bastian}, {Baum}, {Baur}, {Bay}, {Beatty}, {Becker}, {Becker Tjus}, {BenZvi}, {Berley}, {Bernardini}, {Besson}, {Binder}, {Bindig}, {Blaufuss}, {Blot}, {Bohm}, {B{\"o}ser}, {Botner}, {B{\"o}ttcher}, {Bourbeau}, {Bourbeau}, {Bradascio}, {Braun}, {Bron}, {Brostean-Kaiser}, {Burgman}, {Buscher}, {Busse}, {Carver}, {Chen}, {Cheung}, {Chirkin}, {Choi}, {Clark}, {Clark}, {Classen}, {Coleman}, {Collin}, {Conrad}, {Coppin}, {Correa}, {Cowen}, {Cross}, {Dave}, {De Clercq}, {DeLaunay}, {Dembinski}, {Deoskar}, {De Ridder}, {Desiati}, {de Vries}, {de Wasseige}, {de With}, {DeYoung}, {Diaz}, {D{\'\i}az-V{\'e}lez}, {Dujmovic}, {Dunkman}, {Dvorak}, {Eberhardt}, {Ehrhardt}, {Eller}, {Engel}, {Evenson}, {Fahey}, {Fazely}, {Felde}, {Filimonov}, {Finley}, {Fox}, {Franckowiak}, {Friedman}, {Fritz}, {Gaisser}, {Gallagher}, {Ganster}, {Garrappa}, {Gerhardt}, {Ghorbani}, {Glauch}, {Gl{\"u}senkamp}, {Goldschmidt}, {Gonzalez}, {Grant}, {Gr{\'e}goire}, {Griffith}, {Griswold}, {G{\"u}nder}, {G{\"u}nd{\"u}z}, {Haack}, {Hallgren}, {Halliday}, {Halve}, {Halzen}, {Hanson}, {Haungs}, {Hebecker}, {Heereman}, {Heix}, {Helbing}, {Hellauer}, {Henningsen}, {Hickford}, {Hignight}, {Hill}, {Hoffman}, {Hoffmann}, {Hoinka}, {Hokanson-Fasig}, {Hoshina}, {Huang}, {Huber}, {Huber}, {Hultqvist}, {H{\"u}nnefeld}, {Hussain}, {In}, {Iovine}, {Ishihara}, {Jansson}, {Japaridze}, {Jeong}, {Jero}, {Jones}, {Jonske}, {Joppe}, {Kang}, {Kang}, {Kappes}, {Kappesser}, {Karg}, {Karl}, {Karle}, {Katz}, {Kauer}, {Kellermann}, {Kelley}, {Kheirandish}, {Kim}, {Kintscher}, {Kiryluk}, {Kittler}, {Klein}, {Koirala}, {Kolanoski}, {K{\"o}pke}, {Kopper}, {Kopper}, {Koskinen}, {Kowalski}, {Krings}, {Kr{\"u}ckl}, {Kulacz}, {Kurahashi}, {Kyriacou}, {Lanfranchi}, {Larson}, {Lauber}, {Lazar}, {Leonard}, {Leszczy{\'n}ska}, {Liu}, {Lohfink}, {Lozano Mariscal}, {Lu}, {Lucarelli}, {Ludwig}, {L{\"u}nemann}, {Luszczak}, {Lyu}, {Ma}, {Madsen}, {Maggi}, {Mahn}, {Makino}, {Mallik}, {Mallot}, {Mancina}, {Mari{\c{s}}}, {Maruyama}, {Mase}, {Maunu}, {McNally}, {Meagher}, {Medici}, {Medina}, {Meier}, {Meighen-Berger}, {Merino}, {Meures}, {Micallef}, {Mockler}, {Moment{\'e}}, {Montaruli}, {Moore}, {Morse}, {Moulai}, {Muth}, {Nagai}, {Naumann}, {Neer}, {Nguy{\^e}n}, {Niederhausen}, {Nisa}, {Nowicki}, {Nygren}, {Obertacke Pollmann}, {Oehler}, {Olivas}, {O'Murchadha}, {O'Sullivan}, {Palczewski}, {Pandya}, {Pankova}, {Park}, {Peiffer}, {P{\'e}rez de los Heros}, {Philippen}, {Pieloth}, {Pieper}, {Pinat}, {Pizzuto}, {Plum}, {Porcelli}, {Price}, {Przybylski}, {Raab}, {Raissi}, {Rameez}, {Rauch}, {Rawlins}, {Rea}, {Rehman}, {Reimann}, {Relethford}, {Renschler}, {Renzi}, {Resconi}, {Rhode}, {Richman}, {Robertson}, {Rongen}, {Rott}, {Ruhe}, {Ryckbosch}, {Rysewyk Cantu}, {Safa}, {Sanchez Herrera}, {Sandrock}, {Sandroos}, {Santander}, {Sarkar}, {Sarkar}, {Satalecka}, {Schaufel}, {Schieler}, {Schlunder}, {Schmidt}, {Schneider}, {Schneider}, {Schr{\"o}der}, {Schumacher}, {Sclafani}, {Seckel}, {Seunarine}, {Shefali}, {Silva}, {Snihur}, {Soedingrekso}, {Soldin}, {Song}, {Spiczak}, {Spiering}, {Stachurska}, {Stamatikos}, {Stanev}, {Stein}, {Stettner}, {Steuer}, {Stezelberger}, {Stokstad}, {St{\"o}{\ss}l}, {Strotjohann}, {St{\"u}rwald}, {Stuttard}, {Sullivan}, {Taboada}, {Tenholt}, {Ter-Antonyan}, {Terliuk}, {Tilav}, {Tollefson}, {Tomankova}, {T{\"o}nnis}, {Toscano}, {Tosi}, {Trettin}, {Tselengidou}, {Tung}, {Turcati}, {Turcotte}, {Turley}, {Ty}, {Unger}, {Unland Elorrieta}, {Usner}, {Vandenbroucke}, {Van Driessche}, {van Eijk}, {van Eijndhoven}, {van Santen}, {Verpoest}, {Vraeghe}, {Walck}, {Wallace}, {Wallraff}, {Wandkowsky}, {Watson}, {Weaver}, {Weindl}, {Weiss}, {Weldert}, {Wendt}, {Werthebach}, {Whelan}, {Whitehorn}, {Wiebe}, {Wiebusch}, {Wille}, {Williams}, {Wills}, {Wolf}, {Wood}, {Wood}, {Woschnagg}, {Wrede}, {Xu}, {Xu}, {Xu}, {Yanez}, {Yodh}, {Yoshida}, {Yuan}, {Z{\"o}cklein}, \& {IceCube Collaboration}}]{Aartsen2020ApJ...898..117A}
---. 2020, \apj, 898, 117, \dodoi{10.3847/1538-4357/ab9fa0}

\bibitem[{{Aartsen} {et~al.}(2021){Aartsen}, {Abbasi}, {Ackermann}, {Adams}, {Aguilar}, {Ahlers}, {Ahrens}, {Alispach}, {Allison}, {Amin}, {Andeen}, {Anderson}, {Ansseau}, {Anton}, {Arg{\"u}elles}, {Arlen}, {Auffenberg}, {Axani}, {Bagherpour}, {Bai}, {Balagopal V}, {Barbano}, {Bartos}, {Bastian}, {Basu}, {Baum}, {Baur}, {Bay}, {Beatty}, {Becker}, {Tjus}, {BenZvi}, {Berley}, {Bernardini}, {Besson}, {Binder}, {Bindig}, {Blaufuss}, {Blot}, {Bohm}, {Bohmer}, {B{\"o}ser}, {Botner}, {B{\"o}ttcher}, {Bourbeau}, {Bourbeau}, {Bradascio}, {Braun}, {Bron}, {Brostean-Kaiser}, {Burgman}, {Burley}, {Buscher}, {Busse}, {Bustamante}, {Campana}, {Carnie-Bronca}, {Carver}, {Chen}, {Chen}, {Cheung}, {Chirkin}, {Choi}, {Clark}, {Clark}, {Classen}, {Coleman}, {Collin}, {Connolly}, {Conrad}, {Coppin}, {Correa}, {Cowen}, {Cross}, {Dave}, {Deaconu}, {De Clercq}, {DeLaunay}, {De Kockere}, {Dembinski}, {Deoskar}, {De Ridder}, {Desai}, {Desiati}, {de Vries}, {de Wasseige}, {de With}, {DeYoung}, {Dharani}, {Diaz}, {D{\'\i}az-V{\'e}lez}, {Dujmovic}, {Dunkman}, {DuVernois}, {Dvorak}, {Ehrhardt}, {Eller}, {Engel}, {Evans}, {Evenson}, {Fahey}, {Farrag}, {Fazely}, {Felde}, {Fienberg}, {Filimonov}, {Finley}, {Fischer}, {Fox}, {Franckowiak}, {Friedman}, {Fritz}, {Gaisser}, {Gallagher}, {Ganster}, {Garcia-Fernandez}, {Garrappa}, {Gartner}, {Gerhard}, {Gernhaeuser}, {Ghadimi}, {Glaser}, {Glauch}, {Gl{\"u}senkamp}, {Goldschmidt}, {Gonzalez}, {Goswami}, {Grant}, {Gr{\'e}goire}, {Griffith}, {Griswold}, {G{\"u}nd{\"u}z}, {Haack}, {Hallgren}, {Halliday}, {Halve}, {Halzen}, {Hanson}, {Hanson}, {Hardin}, {Haugen}, {Haungs}, {Hauser}, {Hebecker}, {Heinen}, {Heix}, {Helbing}, {Hellauer}, {Henningsen}, {Hickford}, {Hignight}, {Hill}, {Hill}, {Hoffman}, {Hoffmann}, {Hoffmann}, {Hoinka}, {Hokanson-Fasig}, {Holzapfel}, {Hoshina}, {Huang}, {Huber}, {Huber}, {Huege}, {Hughes}, {Hultqvist}, {H{\"u}nnefeld}, {Hussain}, {In}, {Iovine}, {Ishihara}, {Jansson}, {Japaridze}, {Jeong}, {Jones}, {Jonske}, {Joppe}, {Kalekin}, {Kang}, {Kang}, {Kang}, {Kappes}, {Kappesser}, {Karg}, {Karl}, {Karle}, {Katori}, {Katz}, {Kauer}, {Keivani}, {Kellermann}, {Kelley}, {Kheirandish}, {Kim}, {Kin}, {Kintscher}, {Kiryluk}, {Kittler}, {Kleifges}, \& {Klein}}]{Aartsen2021JPhG...48f0501A}
{Aartsen}, M.~G., {Abbasi}, R., {Ackermann}, M., {et~al.} 2021, Journal of Physics G Nuclear Physics, 48, 060501, \dodoi{10.1088/1361-6471/abbd48}

\bibitem[{{Abbasi} {et~al.}(2023){Abbasi}, {Ackermann}, {Adams}, {Aguilar}, {Ahlers}, {Ahrens}, {Alameddine}, {Alves}, {Amin}, {Andeen}, {Anderson}, {Anton}, {Arguelles}, {Ashida}, {Athanasiadou}, {Axani}, {Bai}, {Balagopal}, {Barwick}, {Basu}, {Baur}, {Bay}, {Beatty}, {Becker}, {Becker Tjus}, {Beise}, {Bellenghi}, {Benda}, {Benzvi}, {Berley}, {Bernardini}, {Besson}, {Binder}, {Bindig}, {Blaufuss}, {Blot}, {Boddenberg}, {Bontempo}, {Book}, {Borowka}, {Boser}, {Botner}, {Bottcher}, {Bourbeau}, {Bradascio}, {Braun}, {Brinson}, {Bron}, {Brostean-Kaiser}, {Burley}, {Busse}, {Campana}, {Carnie-Bronca}, {Chen}, {Chen}, {Chirkin}, {Choi}, {Clark}, {Clark}, {Classen}, {Coleman}, {Collin}, {Connolly}, {Conrad}, {Coppin}, {Correa}, {Cowen}, {Cross}, {Dappen}, {Dave}, {de Clercq}, {Delaunay}, {Delgado Lopez}, {Dembinski}, {Deoskar}, {Desai}, {Desiati}, {de Vries}, {de Wasseige}, {Deyoung}, {Diaz}, {Diaz-Velez}, {Dittmer}, {Dujmovic}, {Dunkman}, {Duvernois}, {Ehrhardt}, {Eller}, {Engel}, {Erpenbeck}, {Evans}, {Evenson}, {Fan}, {Fazely}, {Fedynitch}, {Feigl}, {Fiedlschuster}, {Fienberg}, {Finley}, {Fischer}, {Fox}, {Franckowiak}, {Friedman}, {Fritz}, {Furst}, {Gaisser}, {Gallagher}, {Ganster}, {Garcia}, {Garrappa}, {Gerhardt}, {Ghadimi}, {Glaser}, {Glauch}, {Glusenkamp}, {Goehlke}, {Goldschmidt}, {Gonzalez}, {Goswami}, {Grant}, {Gregoire}, {Griswold}, {Gunther}, {Gutjahr}, {Haack}, {Hallgren}, {Halliday}, {Halve}, {Halzen}, {Ha}, {Hanson}, {Hardin}, {Harnisch}, {Haungs}, {Helbing}, {Henningsen}, {Hettinger}, {Hickford}, {Hignight}, {Hill}, {Hill}, {Hoffman}, {Hoshina}, {Hou}, {Huang}, {Huber}, {Huber}, {Hultqvist}, {Hunnefeld}, {Hussain}, {Hymon}, {in}, {Iovine}, {Ishihara}, {Jansson}, {Japaridze}, {Jeong}, {Jin}, {Jones}, {Kang}, {Kang}, {Kang}, {Kappes}, {Kappesser}, {Kardum}, {Karg}, {Karl}, {Karle}, {Katz}, {Kauer}, {Kellermann}, {Kelley}, {Kheirandish}, {Kin}, {Kiryluk}, {Klein}, {Kochocki}, {Koirala}, {Kolanoski}, {Kontrimas}, {Kopke}, {Kopper}, {Kopper}, {Koskinen}, {Koundal}, {Kovacevich}, {Kowalski}, {Kozynets}, {Krupczak}, {Kun}, {Kurahashi}, {Lad}, {Lagunas Gualda}, {Lanfranchi}, {Larson}, {Lauber}, {Lazar}, {Lee}, {Leonard}, {Leszczynska}, {Li}, {Lincetto}, {Liu}, {Liubarska}, {Lohfink}, {Lozano Mariscal}, {Lu}, {Lucarelli}, {Ludwig}, {Luszczak}, {Lyu}, {Ma}, {Madsen}, {Mahn}, {Makino}, {Mancina}, {Maris}, {Martinez-Soler}, {Maruyama}, {McCarthy}, {McElroy}, {McNally}, {Mead}, {Meagher}, {Mechbal}, {Medina}, {Meier}, {Meighen-Berger}, {Merckx}, {Micallef}, {Mockler}, {Montaruli}, {Moore}, {Morik}, {Morse}, {Moulai}, {Mukherjee}, {Naab}, {Nagai}, {Nahnhauer}, {Naumann}, {Necker}, {Nguyen}, {Niederhausen}, {Nisa}, {Nowicki}, {Nygren}, {Obertacke Pollmann}, {Oehler}, {Oeyen}, {Olivas}, {O'Sullivan}, {Pandya}, {Pankova}, {Park}, {Parker}, {Paudel}, {Paul}, {Perez de Los Heros}, {Peters}, {Peterson}, {Philippen}, {Pieper}, {Pizzuto}, {Plum}, {Popovych}, {Porcelli}, {Rodriguez}, {Pries}, {Przybylski}, {Raab}, {Rack-Helleis}, {Raissi}, {Rameez}, {Rawlins}, {Rea}, {Rechav}, {Rehman}, {Reichherzer}, {Reimann}, {Renzi}, {Resconi}, {Reusch}, {Rhode}, {Richman}, {Riedel}, {Roberts}, {Robertson}, {Roellinghoff}, {Rongen}, {Rott}, {Ruhe}, {Ryckbosch}, {Rysewyk Cantu}, {Safa}, {Saffer}, {Salazar-Gallegos}, {Sampathkumar}, {Herrera}, {Sandrock}, {Santander}, {Sarkar}, {Sarkar}, {Satalecka}, {Schaufel}, {Schieler}, {Schindler}, {Schmidt}, {Schneider}, {Schneider}, {Schroder}, {Schumacher}, {Schwefer}, {Sclafani}, {Seckel}, {Seunarine}, {Sharma}, {Shefali}, {Shimizu}, {Silva}, {Skrzypek}, {Smithers}, {Snihur}, {Soedingrekso}, {Sogaard}, {Soldin}, {Spannfellner}, {Spiczak}, {Spiering}, {Stamatikos}, {Stanev}, {Stein}, {Stettner}, {Stezelberger}, {Stokstad}, {Sturwald}, {Stuttard}, {Sullivan}, {Taboada}, {Ter-Antonyan}, {Thwaites}, {Tilav}, {Tischbein}, {Tollefson}, {Tonnis}, {Toscano}, {Tosi}, {Trettin}, {Tselengidou}, {Tung}, {Turcati}, {Turcotte}, {Turley}, {Twagirayezu}, {Ty}, {Elorrieta}, {Valtonen-Mattila}, {Vandenbroucke}, {van Eijndhoven}, {Vannerom}, {van Santen}, {Veitch-Michaelis}, {Verpoest}, {Walck}, {Wang}, {Watson}, {Weaver}, {Weigel}, {Weindl}, {Weiss}, {Weldert}, {Wendt}, {Werthebach}, {Weyrauch}, {Whitehorn}, {Wiebusch}, {Willey}, {Williams}, {Wolf}, {Wrede}, {Wulff}, {Xu}, {Yanez}, {Yildizci}, {Yoshida}, {Yu}, {Yuan}, {Zhang}, \& {Zhelnin}}]{Abbasi2023Sci...380.1338I}
{Abbasi}, R., {Ackermann}, M., {Adams}, J., {et~al.} 2023, Science, 380, 1338, \dodoi{10.1126/science.adc9818}

\bibitem[{{Abeysekara} {et~al.}(2019){Abeysekara}, {Albert}, {Alfaro}, {Alvarez}, {{\'A}lvarez}, {Camacho}, {Arceo}, {Arteaga-Vel{\'a}zquez}, {Arunbabu}, {Avila Rojas}, {Ayala Solares}, {Baghmanyan}, {Belmont-Moreno}, {BenZvi}, {Brisbois}, {Caballero-Mora}, {Capistr{\'a}n}, {Carrami{\~n}ana}, {Casanova}, {Cotti}, {Cotzomi}, {Couti{\~n}o de Le{\'o}n}, {De la Fuente}, {de Le{\'o}n}, {Dichiara}, {Dingus}, {DuVernois}, {D{\'\i}az-V{\'e}lez}, {Ellsworth}, {Engel}, {Espinoza}, {Fick}, {Fleischhack}, {Fraija}, {Galv{\'a}n-G{\'a}mez}, {Garc{\'\i}a-Gonz{\'a}lez}, {Garfias}, {Gonz{\'a}lez}, {Goodman}, {Harding}, {Hernandez}, {Hinton}, {Hona}, {Hueyotl-Zahuantitla}, {Hui}, {H{\"u}ntemeyer}, {Iriarte}, {Jardin-Blicq}, {Joshi}, {Kaufmann}, {Kieda}, {Lara}, {Lee}, {Le{\'o}n Vargas}, {Linnemann}, {Longinotti}, {Luis-Raya}, {Lundeen}, {Malone}, {Marinelli}, {Martinez}, {Martinez-Castellanos}, {Mart{\'\i}nez-Castro}, {Mart{\'\i}nez-Huerta}, {Matthews}, {Miranda-Romagnoli}, {Morales-Soto}, {Moreno}, {Mostaf{\'a}}, {Nayerhoda}, {Nellen}, {Newbold}, {Nisa}, {Noriega-Papaqui}, {Peisker}, {P{\'e}rez-P{\'e}rez}, {Pretz}, {Ren}, {Rho}, {Rivi{\`e}re}, {Rosa-Gonz{\'a}lez}, {Rosenberg}, {Ruiz-Velasco}, {Salazar}, {Salesa Greus}, {Sandoval}, {Schneider}, {Schoorlemmer}, {Seglar Arroyo}, {Sinnis}, {Smith}, {Springer}, {Surajbali}, {Tabachnick}, {Tanner}, {Tibolla}, {Tollefson}, {Torres}, {Weisgarber}, {Westerhoff}, {Wood}, {Yapici}, {Zepeda}, {Zhou}, \& {HAWC Collaboration}}]{Abeysekara2019ApJ...881..134A}
{Abeysekara}, A.~U., {Albert}, A., {Alfaro}, R., {et~al.} 2019, \apj, 881, 134, \dodoi{10.3847/1538-4357/ab2f7d}

\bibitem[{{Acciari} {et~al.}(2020){Acciari}, {Ansoldi}, {Antonelli}, {Arbet Engels}, {Baack}, {Babi{\'c}}, {Banerjee}, {Barres de Almeida}, {Barrio}, {Becerra Gonz{\'a}lez}, {Bednarek}, {Bellizzi}, {Bernardini}, {Berti}, {Besenrieder}, {Bhattacharyya}, {Bigongiari}, {Biland}, {Blanch}, {Bonnoli}, {Bo{\v{s}}njak}, {Busetto}, {Carosi}, {Ceribella}, {Chai}, {Chilingaryan}, {Cikota}, {Colak}, {Colin}, {Colombo}, {Contreras}, {Cortina}, {Covino}, {D'Elia}, {da Vela}, {Dazzi}, {de Angelis}, {de Lotto}, {Delfino}, {Delgado}, {Depaoli}, {di Pierro}, {di Venere}, {Do Souto Espi{\~n}eira}, {Dominis Prester}, {Donini}, {Dorner}, {Doro}, {Elsaesser}, {Fallah Ramazani}, {Fattorini}, {Ferrara}, {Fidalgo}, {Foffano}, {Fonseca}, {Font}, {Fruck}, {Fukami}, {Garc{\'\i}a L{\'o}pez}, {Garczarczyk}, {Gasparyan}, {Gaug}, {Giglietto}, {Giordano}, {Godinovi{\'c}}, {Green}, {Guberman}, {Hadasch}, {Hahn}, {Herrera}, {Hoang}, {Hrupec}, {H{\"u}tten}, {Inada}, {Inoue}, {Ishio}, {Iwamura}, {Jouvin}, {Kerszberg}, {Kubo}, {Kushida}, {Lamastra}, {Lelas}, {Leone}, {Lindfors}, {Lombardi}, {Longo}, {L{\'o}pez}, {L{\'o}pez-Coto}, {L{\'o}pez-Oramas}, {Loporchio}, {Machado de Oliveira Fraga}, {Maggio}, {Majumdar}, {Makariev}, {Mallamaci}, {Maneva}, {Manganaro}, {Mannheim}, {Maraschi}, {Mariotti}, {Mart{\'\i}nez}, {Mazin}, {Mi{\'c}anovi{\'c}}, {Miceli}, {Minev}, {Miranda}, {Mirzoyan}, {Molina}, {Moralejo}, {Morcuende}, {Moreno}, {Moretti}, {Munar-Adrover}, {Neustroev}, {Nigro}, {Nilsson}, {Ninci}, {Nishijima}, {Noda}, {Nogu{\'e}s}, {Nozaki}, {Paiano}, {Palacio}, {Palatiello}, {Paneque}, {Paoletti}, {Paredes}, {Pe{\~n}il}, {Peresano}, {Persic}, {Prada Moroni}, {Prandini}, {Puljak}, {Rhode}, {Rib{\'o}}, {Rico}, {Righi}, {Rugliancich}, {Saha}, {Sahakyan}, {Saito}, {Sakurai}, {Satalecka}, {Schmidt}, {Schweizer}, {Sitarek}, {{\v{S}}nidari{\'c}}, {Sobczynska}, {Somero}, {Stamerra}, {Strom}, {Strzys}, {Suda}, {Suri{\'c}}, {Takahashi}, {Tavecchio}, {Temnikov}, {Terzi{\'c}}, {Teshima}, {Torres-Alb{\`a}}, {Tosti}, {Vagelli}, {van Scherpenberg}, {Vanzo}, {Vazquez Acosta}, {Vigorito}, {Vitale}, {Vovk}, {Will}, \& {Zari{\'c}}}]{Acciari2020A&A...635A.158M}
{Acciari}, V.~A., {Ansoldi}, S., {Antonelli}, L.~A., {et~al.} 2020, \aap, 635, A158, \dodoi{10.1051/0004-6361/201936899}

\bibitem[{{Adri{\'a}n-Mart{\'\i}nez} {et~al.}(2016{\natexlab{a}}){Adri{\'a}n-Mart{\'\i}nez}, {Ageron}, {Aharonian}, {Aiello}, {Albert}, {Ameli}, {Anassontzis}, {Andre}, {Androulakis}, {Anghinolfi}, {Anton}, {Ardid}, {Avgitas}, {Barbarino}, {Barbarito}, {Baret}, {Barrios-Mart{\'\i}}, {Belhorma}, {Belias}, {Berbee}, {van den Berg}, {Bertin}, {Beurthey}, {van Beveren}, {Beverini}, {Biagi}, {Biagioni}, {Billault}, {Bond{\`\i}}, {Bormuth}, {Bouhadef}, {Bourlis}, {Bourret}, {Boutonnet}, {Bouwhuis}, {Bozza}, {Bruijn}, {Brunner}, {Buis}, {Busto}, {Cacopardo}, {Caillat}, {Calamai}, {Calvo}, {Capone}, {Caramete}, {Cecchini}, {Celli}, {Champion}, {Cherkaoui El Moursli}, {Cherubini}, {Chiarusi}, {Circella}, {Classen}, {Cocimano}, {Coelho}, {Coleiro}, {Colonges}, {Coniglione}, {Cordelli}, {Cosquer}, {Coyle}, {Creusot}, {Cuttone}, {D'Amico}, {De Bonis}, {De Rosa}, {De Sio}, {Di Capua}, {Di Palma}, {D{\'\i}az Garc{\'\i}a}, {Distefano}, {Donzaud}, {Dornic}, {Dorosti-Hasankiadeh}, {Drakopoulou}, {Drouhin}, {Drury}, {Durocher}, {Eberl}, {Eichie}, {van Eijk}, {El Bojaddaini}, {El Khayati}, {Elsaesser}, {Enzenh{\"o}fer}, {Fassi}, {Favali}, {Fermani}, {Ferrara}, {Filippidis}, {Frascadore}, {Fusco}, {Gal}, {Galat{\`a}}, {Garufi}, {Gay}, {Gebyehu}, {Giordano}, {Gizani}, {Gracia}, {Graf}, {Gr{\'e}goire}, {Grella}, {Habel}, {Hallmann}, {van Haren}, {Harissopulos}, {Heid}, {Heijboer}, {Heine}, {Henry}, {Hern{\'a}ndez-Rey}, {Hevinga}, {Hofest{\"a}dt}, {Hugon}, {Illuminati}, {James}, {Jansweijer}, {Jongen}, {de Jong}, {Kadler}, {Kalekin}, {Kappes}, {Katz}, {Keller}, {Kieft}, {Kie{\ss}ling}, {Koffeman}, {Kooijman}, {Kouchner}, {Kulikovskiy}, {Lahmann}, {Lamare}, {Leisos}, {Leonora}, {Clark}, {Liolios}, {Llorens Alvarez}, {Lo Presti}, {L{\"o}hner}, {Lonardo}, {Lotze}, {Loucatos}, {Maccioni}, {Mannheim}, {Margiotta}, {Marinelli}, {Mari{\c{s}}}, {Markou}, {Mart{\'\i}nez-Mora}, {Martini}, {Mele}, {Melis}, {Michael}, {Migliozzi}, {Migneco}, {Mijakowski}, {Miraglia}, {Mollo}, {Mongelli}, {Morganti}, {Moussa}, {Musico}, {Musumeci}, {Navas}, {Nicolau}, {Olcina}, {Olivetto}, {Orlando}, {Papaikonomou}, {Papaleo}, {P{\u{a}}v{\u{a}}la{\c{s}}}, {Peek}, {Pellegrino}, {Perrina}, {Pfutzner}, {Piattelli}, {Pikounis}, {Poma}, {Popa}, {Pradier}, {Pratolongo}, {P{\"u}hlhofer}, {Pulvirenti}, {Quinn}, {Racca}, {Raffaelli}, {Randazzo}, {Rapidis}, {Razis}, {Real}, {Resvanis}, {Reubelt}, {Riccobene}, {Rossi}, {Rovelli}, {Salda{\~n}a}, {Salvadori}, {Samtleben}, {S{\'a}nchez Garc{\'\i}a}, {S{\'a}nchez Losa}, {Sanguineti}, {Santangelo}, {Santonocito}, {Sapienza}, {Schimmel}, {Schmelling}, {Sciacca}, {Sedita}, {Seitz}, {Sgura}, {Simeone}, {Siotis}, {Sipala}, {Spisso}, {Spurio}, {Stavropoulos}, {Steijger}, {Stellacci}, {Stransky}, {Taiuti}, {Tayalati}, {T{\'e}zier}, {Theraube}, {Thompson}, {Timmer}, {T{\"o}nnis}, {Trasatti}, {Trovato}, {Tsirigotis}, {Tzamarias}, {Tzamariudaki}, {Vallage}, {Van Elewyck}, {Vermeulen}, {Vicini}, {Viola}, {Vivolo}, {Volkert}, {Voulgaris}, {Wiggers}, {Wilms}, {de Wolf}, {Zachariadou}, {Zornoza}, \& {Z{\'u}{\~n}iga}}]{Adrian-Martinez2016JPhG...43h4001A}
{Adri{\'a}n-Mart{\'\i}nez}, S., {Ageron}, M., {Aharonian}, F., {et~al.} 2016{\natexlab{a}}, Journal of Physics G Nuclear Physics, 43, 084001, \dodoi{10.1088/0954-3899/43/8/084001}

\bibitem[{{Adri{\'a}n-Mart{\'\i}nez} {et~al.}(2016{\natexlab{b}}){Adri{\'a}n-Mart{\'\i}nez}, {Albert}, {Andr{\'e}}, {Anghinolfi}, {Anton}, {Ardid}, {Aubert}, {Avgitas}, {Baret}, {Barrios-Mart{\'\i}}, {Basa}, {Bertin}, {Biagi}, {Bormuth}, {Bouwhuis}, {Bruijn}, {Brunner}, {Busto}, {Capone}, {Caramete}, {Carr}, {Celli}, {Chiarusi}, {Circella}, {Coleiro}, {Coniglione}, {Costantini}, {Coyle}, {Creusot}, {Deschamps}, {De Bonis}, {Distefano}, {Donzaud}, {Dornic}, {Drouhin}, {Eberl}, {El Bojaddaini}, {Els{\"a}sser}, {Enzenh{\"o}fer}, {Fehn}, {Felis}, {Fusco}, {Galat{\`a}}, {Gay}, {Gei{\ss}els{\"o}der}, {Geyer}, {Giordano}, {Gleixner}, {Glotin}, {Gracia-Ruiz}, {Graf}, {Hallmann}, {van Haren}, {Heijboer}, {Hello}, {Hern{\'a}ndez-Rey}, {H{\"o}{\ss}l}, {Hofest{\"a}dt}, {Hugon}, {Illuminati}, {James}, {de Jong}, {Kadler}, {Kalekin}, {Katz}, {Kie{\ss}ling}, {Kouchner}, {Kreter}, {Kreykenbohm}, {Kulikovskiy}, {Lachaud}, {Lahmann}, {Lef{\`e}vre}, {Leonora}, {Loucatos}, {Marcelin}, {Margiotta}, {Marinelli}, {Mart{\'\i}nez-Mora}, {Mathieu}, {Michael}, {Migliozzi}, {Moussa}, {Mueller}, {Nezri}, {P{\u{a}}v{\u{a}}la{\c{s}}}, {Pellegrino}, {Perrina}, {Piattelli}, {Popa}, {Pradier}, {Racca}, {Riccobene}, {Roensch}, {Salda{\~n}a}, {Samtleben}, {S{\'a}nchez-Losa}, {Sanguineti}, {Sapienza}, {Schnabel}, {Sch{\"u}ssler}, {Seitz}, {Sieger}, {Spurio}, {Stolarczyk}, {Taiuti}, {Trovato}, {Tselengidou}, {Turpin}, {T{\"o}nnis}, {Vallage}, {Vall{\'e}e}, {Van Elewyck}, {Visser}, {Vivolo}, {Wagner}, {Wilms}, {Zornoza}, \& {Z{\'u}{\~n}iga}}]{Adrian-Martinez2016PhLB..760..143A}
{Adri{\'a}n-Mart{\'\i}nez}, S., {Albert}, A., {Andr{\'e}}, M., {et~al.} 2016{\natexlab{b}}, Physics Letters B, 760, 143, \dodoi{10.1016/j.physletb.2016.06.051}

\bibitem[{{Ageron} {et~al.}(2011){Ageron}, {Aguilar}, {Al Samarai}, {Albert}, {Ameli}, {Andr{\'e}}, {Anghinolfi}, {Anton}, {Anvar}, {Ardid}, {Arnaud}, {Aslanides}, {Assis Jesus}, {Astraatmadja}, {Aubert}, {Auer}, {Barbarito}, {Baret}, {Basa}, {Bazzotti}, {Becherini}, {Beltramelli}, {Bersani}, {Bertin}, {Beurthey}, {Biagi}, {Bigongiari}, {Billault}, {Blaes}, {Bogazzi}, {de Botton}, {Bou-Cabo}, {Boudahef}, {Bouwhuis}, {Brown}, {Brunner}, {Busto}, {Caillat}, {Calzas}, {Camarena}, {Capone}, {Caponetto}, {C{\^a}rloganu}, {Carminati}, {Carmona}, {Carr}, {Carton}, {Cassano}, {Castorina}, {Cecchini}, {Ceres}, {Chaleil}, {Charvis}, {Chauchot}, {Chiarusi}, {Circella}, {Comp{\`e}re}, {Coniglione}, {Coppolani}, {Cosquer}, {Costantini}, {Cottini}, {Coyle}, {Cuneo}, {Curtil}, {D'Amato}, {Damy}, {van Dantzig}, {de Bonis}, {Decock}, {Decowski}, {Dekeyser}, {Delagnes}, {Desages-Ardellier}, {Deschamps}, {Destelle}, {di Maria}, {Dinkespiler}, {Distefano}, {Dominique}, {Donzaud}, {Dornic}, {Dorosti}, {Drogou}, {Drouhin}, {Druillole}, {Durand}, {Durand}, {Eberl}, {Emanuele}, {Engelen}, {Ernenwein}, {Escoffier}, {Falchini}, {Favard}, {Fehr}, {Feinstein}, {Ferri}, {Ferry}, {Fiorello}, {Flaminio}, {Folger}, {Fritsch}, {Fuda}, {Galat{\'a}}, {Galeotti}, {Gay}, {Gensolen}, {Giacomelli}, {Gojak}, {G{\'o}mez-Gonz{\'a}lez}, {Goret}, {Graf}, {Guillard}, {Halladjian}, {Hallewell}, {van Haren}, {Hartmann}, {Heijboer}, {Heine}, {Hello}, {Henry}, {Hern{\'a}ndez-Rey}, {Herold}, {H{\"o}{\ss}l}, {Hogenbirk}, {Hsu}, {Hubbard}, {Jaquet}, {Jaspers}, {de Jong}, {Jourde}, {Kadler}, {Kalantar-Nayestanaki}, {Kalekin}, {Kappes}, {Karg}, {Karkar}, {Karolak}, {Katz}, {Keller}, {Kestener}, {Kok}, {Kok}, {Kooijman}, {Kopper}, {Kouchner}, {Kretschmer}, {Kruijer}, {Kuch}, {Kulikovskiy}, {Lachartre}, {Lafoux}, {Lagier}, {Lahmann}, {Lahonde-Hamdoun}, {Lamare}, {Lambard}, {Languillat}, {Larosa}, {Lavalle}, {Le Guen}, {Le Provost}, {Levansuu}, {Lef{\`e}vre}, {Legou}, {Lelaizant}, {L{\'e}v{\'e}que}, {Lim}, {Lo Presti}, {Loehner}, {Loucatos}, {Louis}, {Lucarelli}, {Lyashuk}, {Magnier}, {Mangano}, {Marcel}, {Marcelin}, {Margiotta}, {Martinez-Mora}, {Masullo}, {Maz{\'e}as}, {Mazure}, {Meli}, {Melissas}, {Migneco}, {Mongelli}, {Montaruli}, {Morganti}, {Moscoso}, {Motz}, {Musumeci}, {Naumann}, {Naumann-Godo}, {Neff}, {Niess}, {Nooren}, {Oberski}, {Olivetto}, {Palanque-Delabrouille}, {Palioselitis}, {Papaleo}, {P{\u{a}}v{\u{a}}la{\c{s}}}, {Payet}, {Payre}, {Peek}, {Petrovic}, {Piattelli}, {Picot-Clemente}, {Picq}, {Piret}, {Poinsignon}, {Popa}, {Pradier}, {Presani}, {Prono}, {Racca}, {Raia}, {van Randwijk}, {Real}, {Reed}, {R{\'e}thor{\'e}}, {Rewiersma}, {Riccobene}, {Richardt}, {Richter}, {Ricol}, {Rigaud}, {Roca}, {Roensch}, {Rolin}, {Rostovtsev}, {Rottura}, {Roux}, {Rujoiu}, {Ruppi}, {Russo}, {Salesa}, {Salomon}, {Sapienza}, {Schmitt}, {Sch{\"o}ck}, {Schuller}, {Sch{\"u}ssler}, {Sciliberto}, {Shanidze}, {Shirokov}, {Simeone}, {Sottoriva}, {Spies}, {Spona}, {Spurio}, {Steijger}, {Stolarczyk}, {Streeb}, {Sulak}, {Taiuti}, {Tamburini}, {Tao}, {Tasca}, {Terreni}, {Tezier}, {Toscano}, {Urbano}, {Valdy}, {Vallage}, {van Elewyck}, {Vannoni}, {Vecchi}, {Venekamp}, {Verlaat}, {Vernin}, {Virique}, {de Vries}, {van Wijk}, {Wijnker}, {Wobbe}, {de Wolf}, {Yakovenko}, {Yepes}, {Zaborov}, {Zaccone}, {Zornoza}, \& {Z{\'u}{\~n}iga}}]{Ageron2011NIMPA.656...11A}
{Ageron}, M., {Aguilar}, J.~A., {Al Samarai}, I., {et~al.} 2011, Nuclear Instruments and Methods in Physics Research A, 656, 11, \dodoi{10.1016/j.nima.2011.06.103}

\bibitem[{{Agostini} {et~al.}(2020){Agostini}, {B{\"o}hmer}, {Bosma}, {Clark}, {Danninger}, {Fruck}, {Gernh{\"a}user}, {G{\"a}rtner}, {Grant}, {Henningsen}, {Holzapfel}, {Huber}, {Jenkyns}, {Krauss}, {Krings}, {Kopper}, {Leism{\"u}ller}, {Leys}, {Macoun}, {Meighen-Berger}, {Michel}, {Moore}, {Morley}, {Padovani}, {Papp}, {Pirenne}, {Qiu}, {Rea}, {Resconi}, {Round}, {Ruskey}, {Spannfellner}, {Traxler}, {Turcati}, \& {Yanez}}]{Agostini2020NatAs...4..913A}
{Agostini}, M., {B{\"o}hmer}, M., {Bosma}, J., {et~al.} 2020, Nature Astronomy, 4, 913, \dodoi{10.1038/s41550-020-1182-4}

\bibitem[{{Aharonian} {et~al.}(2004){Aharonian}, {Akhperjanian}, {Beilicke}, {Bernl{\"o}hr}, {B{\"o}rst}, {Bojahr}, {Bolz}, {Coarasa}, {Contreras}, {Cortina}, {Denninghoff}, {Fonseca}, {Girma}, {G{\"o}tting}, {Heinzelmann}, {Hermann}, {Heusler}, {Hofmann}, {Horns}, {Jung}, {Kankanyan}, {Kestel}, {Kohnle}, {Konopelko}, {Kranich}, {Lampeitl}, {Lopez}, {Lorenz}, {Lucarelli}, {Mang}, {Mazin}, {Meyer}, {Mirzoyan}, {Moralejo}, {O{\~n}a-Wilhelmi}, {Panter}, {Plyasheshnikov}, {P{\"u}hlhofer}, {de los Reyes}, {Rhode}, {Ripken}, {Rowell}, {Sahakian}, {Samorski}, {Schilling}, {Siems}, {Sobzynska}, {Stamm}, {Tluczykont}, {Vitale}, {V{\"o}lk}, {Wiedner}, \& {Wittek}}]{Aharonian2004ApJ...614..897A}
{Aharonian}, F., {Akhperjanian}, A., {Beilicke}, M., {et~al.} 2004, \apj, 614, 897, \dodoi{10.1086/423931}

\bibitem[{{Aharonian} {et~al.}(2024){Aharonian}, {Ait Benkhali}, {Aschersleben}, {Ashkar}, {Backes}, {Baktash}, {Barbosa Martins}, {Batzofin}, {Becherini}, {Berge}, {Bernl{\"o}hr}, {Bi}, {B{\"o}ttcher}, {Boisson}, {Bolmont}, {de Bony de Lavergne}, {Borowska}, {Bradascio}, {Breuhaus}, {Brose}, {Brown}, {Brun}, {Bruno}, {Bulik}, {Burger-Scheidlin}, {Bylund}, {Caroff}, {Casanova}, {Cecil}, {Celic}, {Cerruti}, {Chambery}, {Chand}, {Chandra}, {Chen}, {Chibueze}, {Chibueze}, {Cotter}, {Cristofari}, {Devin}, {Djannati-Ata{\"\i}}, {Djuvsland}, {Dmytriiev}, {Einecke}, {Ernenwein}, {Fegan}, {Feijen}, {Filipovi{\'c}}, {Fontaine}, {F{\"u}{\ss}ling}, {Funk}, {Gabici}, {Gallant}, {Giavitto}, {Glawion}, {Glicenstein}, {Glombitza}, {Goswami}, {Grolleron}, {Grondin}, {Haerer}, {Hinton}, {Hofmann}, {Holch}, {Holler}, {Horns}, {Jamrozy}, {Jankowsky}, {Joshi}, {Kasai}, {Katarzy{\'n}ski}, {Khatoon}, {Kh{\'e}lifi}, {Klu{\'z}niak}, {Komin}, {Kosack}, {Kostunin}, {Kundu}, {Lang}, {Le Stum}, {Leitl}, {Lemi{\`e}re}, {Lemoine-Goumard}, {Lenain}, {Leuschner}, {Luashvili}, {Mackey}, {Malyshev}, {Malyshev}, {Marandon}, {Marinos}, {Mart{\'\i}-Devesa}, {Marx}, {Mehta}, {Meyer}, {Mitchell}, {Moderski}, {Mohrmann}, {Montanari}, {Moulin}, {Murach}, {de Naurois}, {Niemiec}, {O'Brien}, {Ohm}, {Olivera-Nieto}, {de Ona Wilhelmi}, {Ostrowski}, {Panny}, {Panter}, {Parsons}, {Peron}, {Prokhorov}, {P{\"u}hlhofer}, {Punch}, {Quirrenbach}, {Regeard}, {Reichherzer}, {Reimer}, {Reimer}, {Ren}, {Renaud}, {Reville}, {Rieger}, {Roellinghoff}, {Rudak}, {Sahakian}, {Salzmann}, {Sasaki}, {Sch{\"u}ssler}, {Schutte}, {Shapopi}, {Specovius}, {Spencer}, {Stawarz}, {Steenkamp}, {Steinmassl}, {Steppa}, {Streil}, {Sushch}, {Suzuki}, {Takahashi}, {Tanaka}, {Terrier}, {Tluczykont}, {Tsuji}, {Unbehaun}, {van Eldik}, {Vecchi}, {Veh}, {Venter}, {Vink}, {Wach}, {Wagner}, {Wierzcholska}, {Zacharias}, {Zargaryan}, {Zdziarski}, {Zech}, {Zouari}, {{\.Z}ywucka}, \& {Harding}}]{Aharonian2024A&A...686A.308A}
{Aharonian}, F., {Ait Benkhali}, F., {Aschersleben}, J., {et~al.} 2024, \aap, 686, A308, \dodoi{10.1051/0004-6361/202348651}

\bibitem[{{Albert} {et~al.}(2017){Albert}, {Andr{\'e}}, {Anghinolfi}, {Anton}, {Ardid}, {Aubert}, {Avgitas}, {Baret}, {Barrios-Mart{\'\i}}, {Basa}, {Belhorma}, {Bertin}, {Biagi}, {Bormuth}, {Bourret}, {Bouwhuis}, {Bruijn}, {Brunner}, {Busto}, {Capone}, {Caramete}, {Carr}, {Celli}, {Cherkaoui El Moursli}, {Chiarusi}, {Circella}, {Coelho}, {Coleiro}, {Coniglione}, {Costantini}, {Coyle}, {Creusot}, {D{\'\i}az}, {Deschamps}, {de Bonis}, {Distefano}, {di Palma}, {Domi}, {Donzaud}, {Dornic}, {Drouhin}, {Eberl}, {El Bojaddaini}, {El Khayati}, {Els{\"a}sser}, {Enzenh{\"o}fer}, {Ettahiri}, {Fassi}, {Felis}, {Fusco}, {Galat{\`a}}, {Gay}, {Giordano}, {Glotin}, {Gr{\'e}goire}, {Gracia Ruiz}, {Graf}, {Hallmann}, {van Haren}, {Heijboer}, {Hello}, {Hern{\'a}ndez-Rey}, {H{\"o}{\ss}l}, {Hofest{\"a}dt}, {Hugon}, {Illuminati}, {James}, {de Jong}, {Jongen}, {Kadler}, {Kalekin}, {Katz}, {Kie{\ss}ling}, {Kouchner}, {Kreter}, {Kreykenbohm}, {Kulikovskiy}, {Lachaud}, {Lahmann}, {Lef{\`e}vre}, {Leonora}, {Lotze}, {Loucatos}, {Marcelin}, {Margiotta}, {Marinelli}, {Mart{\'\i}nez-Mora}, {Mele}, {Melis}, {Michael}, {Migliozzi}, {Moussa}, {Navas}, {Nezri}, {Organokov}, {P{\v{a}}v{\v{a}}la{\c{s}}}, {Pellegrino}, {Perrina}, {Piattelli}, {Popa}, {Pradier}, {Quinn}, {Racca}, {Riccobene}, {S{\'a}nchez-Losa}, {Salda{\~n}a}, {Salvadori}, {Samtleben}, {Sanguineti}, {Sapienza}, {Sch{\"u}ssler}, {Sieger}, {Spurio}, {Stolarczyk}, {Taiuti}, {Tayalati}, {Trovato}, {Turpin}, {T{\"o}nnis}, {Vallage}, {van Elewyck}, {Versari}, {Vivolo}, {Vizzoca}, {Wilms}, {Zornoza}, {Z{\'u}{\~n}iga}, {Gaggero}, {Grasso}, \& {ANTARES Collaboration}}]{Albert2017PhRvD..96f2001A}
{Albert}, A., {Andr{\'e}}, M., {Anghinolfi}, M., {et~al.} 2017, \prd, 96, 062001, \dodoi{10.1103/PhysRevD.96.062001}

\bibitem[{{Albert} {et~al.}(2018){Albert}, {Andr{\'e}}, {Anghinolfi}, {Ardid}, {Aubert}, {Aublin}, {Avgitas}, {Baret}, {Barrios-Mart{\'\i}}, {Basa}, {Belhorma}, {Bertin}, {Biagi}, {Bormuth}, {Boumaaza}, {Bourret}, {Bouwhuis}, {Br{\^a}nza{\c{s}}}, {Bruijn}, {Brunner}, {Busto}, {Capone}, {Caramete}, {Carr}, {Celli}, {Chabab}, {Cherkaoui El Moursli}, {Chiarusi}, {Circella}, {Coelho}, {Coleiro}, {Colomer}, {Coniglione}, {Costantini}, {Coyle}, {Creusot}, {D{\'\i}az}, {Deschamps}, {Distefano}, {Di Palma}, {Domi}, {Donzaud}, {Dornic}, {Drouhin}, {Eberl}, {El Bojaddaini}, {El Khayati}, {Els{\"a}sser}, {Enzenh{\"o}fer}, {Ettahiri}, {Fassi}, {Felis}, {Fermani}, {Ferrara}, {Fusco}, {Gay}, {Glotin}, {Gr{\'e}goire}, {Ruiz}, {Graf}, {Hallmann}, {van Haren}, {Heijboer}, {Hello}, {Hern{\'a}ndez-Rey}, {H{\"o}{\ss}l}, {Hofest{\"a}dt}, {Illuminati}, {James}, {de Jong}, {Jongen}, {Kadler}, {Kalekin}, {Katz}, {Khan-Chowdhury}, {Kouchner}, {Kreter}, {Kreykenbohm}, {Kulikovskiy}, {Lachaud}, {Lahmann}, {Lef{\`e}vre}, {Leonora}, {Levi}, {Lotze}, {Loucatos}, {Marcelin}, {Margiotta}, {Marinelli}, {Mart{\'\i}nez-Mora}, {Mele}, {Melis}, {Migliozzi}, {Moussa}, {Navas}, {Nezri}, {Nu{\~n}ez}, {Organokov}, {P{\u{a}}v{\u{a}}la{\c{s}}}, {Pellegrino}, {Piattelli}, {Popa}, {Pradier}, {Quinn}, {Racca}, {Randazzo}, {Riccobene}, {S{\'a}nchez-Losa}, {Salda{\~n}a}, {Salvadori}, {Samtleben}, {Sanguineti}, {Sapienza}, {Sch{\"u}ssler}, {Spurio}, {Stolarczyk}, {Taiuti}, {Tayalati}, {Trovato}, {Vallage}, {Van Elewyck}, {Versari}, {Vivolo}, {Wilms}, {Zaborov}, {Zornoza}, {Z{\'u}{\~n}iga}, {ANTARES Collaboration}, {Aartsen}, {Ackermann}, {Adams}, {Aguilar}, {Ahlers}, {Ahrens}, {Samarai}, {Altmann}, {Andeen}, {Anderson}, {Ansseau}, {Anton}, {Arg{\"u}elles}, {Auffenberg}, {Axani}, {Backes}, {Bagherpour}, {Bai}, {Barbano}, {Barron}, {Barwick}, {Baum}, {Bay}, {Beatty}, {Becker Tjus}, {Becker}, {BenZvi}, {Berley}, {Bernardini}, {Besson}, {Binder}, {Bindig}, {Blaufuss}, {Blot}, {Bohm}, {B{\"o}rner}, {Bos}, {B{\"o}ser}, {Botner}, {Bourbeau}, {Bourbeau}, {Bradascio}, {Braun}, {Brenzke}, {Bretz}, {Bron}, {Brostean-Kaiser}, {Burgman}, {Busse}, {Carver}, {Cheung}, {Chirkin}, {Christov}, {Clark}, {Classen}, {Collin}, {Conrad}, {Coppin}, {Correa}, {Cowen}, {Cross}, {Dave}, {Day}, {de Andr{\'e}}, {De Clercq}, {DeLaunay}, {Dembinski}, {Deoskar}, {De Ridder}, {Desiati}, {de Vries}, \& {de Wasseige}}]{Albert2018ApJ...868L..20A}
---. 2018, \apjl, 868, L20, \dodoi{10.3847/2041-8213/aaeecf}

\bibitem[{{Albert} {et~al.}(2023){Albert}, {Alves}, {Andr{\'e}}, {Ardid}, {Ardid}, {Aubert}, {Aublin}, {Baret}, {Basa}, {Becherini}, {Belhorma}, {Bendahman}, {Benfenati}, {Bertin}, {Biagi}, {Bissinger}, {Boumaaza}, {Bouta}, {Bouwhuis}, {Br{\^a}nza{\c{s}}}, {Bruijn}, {Brunner}, {Busto}, {Caiffi}, {Calvo}, {Campion}, {Capone}, {Caramete}, {Carenini}, {Carr}, {Carretero}, {Celli}, {Cerisy}, {Chabab}, {Chau}, {Cherkaoui El Moursli}, {Chiarusi}, {Circella}, {Coelho}, {Coleiro}, {Coniglione}, {Coyle}, {Creusot}, {D{\'\i}az}, {de Martino}, {Distefano}, {di Palma}, {Domi}, {Donzaud}, {Dornic}, {Drouhin}, {Eberl}, {van Eeden}, {van Eijk}, {El Hedri}, {El Khayati}, {Enzenh{\"o}fer}, {Fasano}, {Fermani}, {Ferrara}, {Filippini}, {Fusco}, {Gagliardini}, {Garc{\'\i}a}, {Gatius Oliver}, {Gay}, {Gei{\ss}elbrecht}, {Glotin}, {Gozzini}, {Gracia Ruiz}, {Graf}, {Guidi}, {Haegel}, {Hallmann}, {van Haren}, {Heijboer}, {Hello}, {Hern{\'a}ndez-Rey}, {H{\"o}{\ss}l}, {Hofest{\"a}dt}, {Huang}, {Illuminati}, {James}, {Jisse-Jung}, {de Jong}, {de Jong}, {Kadler}, {Kalekin}, {Katz}, {Kouchner}, {Kreykenbohm}, {Kulikovskiy}, {Lahmann}, {Lamoureux}, {Lazo}, {Lef{\`e}vre}, {Leonora}, {Levi}, {Le Stum}, {Lopez-Coto}, {Loucatos}, {Maderer}, {Manczak}, {Marcelin}, {Margiotta}, {Marinelli}, {Mart{\'\i}nez-Mora}, {Migliozzi}, {Moussa}, {Muller}, {Nauta}, {Navas}, {Neronov}, {Nezri}, {{\'O} Fearraigh}, {P{\u{a}}un}, {P{\u{a}}v{\u{a}}la{\c{s}}}, {Perrin-Terrin}, {Pestel}, {Piattelli}, {Poir{\`e}}, {Popa}, {Pradier}, {Randazzo}, {Real}, {Reck}, {Riccobene}, {Romanov}, {S{\'a}nchez-Losa}, {Saina}, {Salesa Greus}, {Samtleben}, {Sanguineti}, {Sapienza}, {Savchenko}, {Schnabel}, {Schumann}, {Sch{\"u}ssler}, {Seneca}, {Spurio}, {Stolarczyk}, {Taiuti}, {Tayalati}, {Tingay}, {Vallage}, {Vannoye}, {van Elewyck}, {Viola}, {Vivolo}, {Wilms}, {Zavatarelli}, {Zegarelli}, {Zornoza}, {Z{\'u}{\~n}iga}, \& {ANTARES Collaboration}}]{Albert2023PhLB..84137951A}
{Albert}, A., {Alves}, S., {Andr{\'e}}, M., {et~al.} 2023, Physics Letters B, 841, 137951, \dodoi{10.1016/j.physletb.2023.137951}

\bibitem[{{Aleksi{\'c}} {et~al.}(2015){Aleksi{\'c}}, {Ansoldi}, {Antonelli}, {Antoranz}, {Babic}, {Bangale}, {Barrio}, {Becerra Gonz{\'a}lez}, {Bednarek}, {Bernardini}, {Biasuzzi}, {Biland}, {Blanch}, {Bonnefoy}, {Bonnoli}, {Borracci}, {Bretz}, {Carmona}, {Carosi}, {Colin}, {Colombo}, {Contreras}, {Cortina}, {Covino}, {Da Vela}, {Dazzi}, {De Angelis}, {De Caneva}, {De Lotto}, {de O{\~n}a Wilhelmi}, {Delgado Mendez}, {Doert}, {Dominis Prester}, {Dorner}, {Doro}, {Einecke}, {Eisenacher}, {Elsaesser}, {Fonseca}, {Font}, {Frantzen}, {Fruck}, {Galindo}, {Garc{\'\i}a L{\'o}pez}, {Garczarczyk}, {Garrido Terrats}, {Gaug}, {Godinovi{\'c}}, {Gonz{\'a}lez Mu{\~n}oz}, {Gozzini}, {Hadasch}, {Hanabata}, {Hayashida}, {Herrera}, {Hildebrand}, {Hose}, {Hrupec}, {Idec}, {Kadenius}, {Kellermann}, {Kodani}, {Konno}, {Krause}, {Kubo}, {Kushida}, {La Barbera}, {Lelas}, {Lewandowska}, {Lindfors}, {Lombardi}, {L{\'o}pez}, {L{\'o}pez-Coto}, {L{\'o}pez-Oramas}, {Lorenz}, {Lozano}, {Makariev}, {Mallot}, {Maneva}, {Mankuzhiyil}, {Mannheim}, {Maraschi}, {Marcote}, {Mariotti}, {Mart{\'\i}nez}, {Mazin}, {Menzel}, {Miranda}, {Mirzoyan}, {Moralejo}, {Munar-Adrover}, {Nakajima}, {Niedzwiecki}, {Nilsson}, {Nishijima}, {Noda}, {Nowak}, {Orito}, {Overkemping}, {Paiano}, {Palatiello}, {Paneque}, {Paoletti}, {Paredes}, {Paredes-Fortuny}, {Persic}, {Prada Moroni}, {Prandini}, {Preziuso}, {Puljak}, {Reinthal}, {Rhode}, {Rib{\'o}}, {Rico}, {Rodriguez Garcia}, {R{\"u}gamer}, {Saggion}, {Saito}, {Saito}, {Satalecka}, {Scalzotto}, {Scapin}, {Schultz}, {Schweizer}, {Shore}, {Sillanp{\"a}{\"a}}, {Sitarek}, {Snidaric}, {Sobczynska}, {Spanier}, {Stamatescu}, {Stamerra}, {Steinbring}, {Storz}, {Strzys}, {Takalo}, {Takami}, {Tavecchio}, {Temnikov}, {Terzi{\'c}}, {Tescaro}, {Teshima}, {Thaele}, {Tibolla}, {Torres}, {Toyama}, {Treves}, {Uellenbeck}, {Vogler}, {Wagner}, {Zanin}, {Horns}, {Mart{\'\i}n}, \& {Meyer}}]{Aleksic2015JHEAp...5...30A}
{Aleksi{\'c}}, J., {Ansoldi}, S., {Antonelli}, L.~A., {et~al.} 2015, Journal of High Energy Astrophysics, 5, 30, \dodoi{10.1016/j.jheap.2015.01.002}

\bibitem[{{Allakhverdyan} {et~al.}(2024){Allakhverdyan}, {Avrorin}, {Avrorin}, {Aynutdinov}, {Barda{\v{c}}ov{\'a}}, {Belolaptikov}, {Bondarev}, {Borina}, {Budnev}, {Chadymov}, {Chepurnov}, {Dik}, {Domogatsky}, {Doroshenko}, {Dvornick{\'y}}, {Dyachok}, {Dzhilkibaev}, {Eckerov{\'a}}, {Elzhov}, {Fomin}, {Gafarov}, {Golubkov}, {Gorshkov}, {Gress}, {Kebkal}, {Kebkal}, {Kharuk}, {Khramov}, {Kleimenov}, {Kolbin}, {Koligaev}, {Konischev}, {Korobchenko}, {Koshechkin}, {Kozhin}, {Kruglov}, {Kulepov}, {Kulikov}, {Lemeshev}, {Mirgazov}, {Naumov}, {Nikolaev}, {Perevalova}, {Petukhov}, {Pliskovsky}, {Rozanov}, {Ryabov}, {Safronov}, {Shaybonov}, {Shishkin}, {Shirokov}, {{\v{S}}imkovic}, {Sirenko}, {Skurikhin}, {Solovjev}, {Sorokovikov}, {{\v{S}}tekl}, {Stromakov}, {Suvorova}, {Tabolenko}, {Tretjak}, {Ulzutuev}, {Yablokova}, {Zaborov}, {Zavyalov}, {Zvezdov}, {Kovalev}, {Plavin}, {Semikoz}, \& {Troitsky}}]{Allakhverdyan2024arXiv241105608A}
{Allakhverdyan}, V.~A., {Avrorin}, A.~D., {Avrorin}, A.~V., {et~al.} 2024, arXiv e-prints, arXiv:2411.05608, \dodoi{10.48550/arXiv.2411.05608}

\bibitem[{{Amato} {et~al.}(2003){Amato}, {Guetta}, \& {Blasi}}]{Amato2003A&A...402..827A}
{Amato}, E., {Guetta}, D., \& {Blasi}, P. 2003, \aap, 402, 827, \dodoi{10.1051/0004-6361:20030279}

\bibitem[{{Amenomori} {et~al.}(2019){Amenomori}, {Bao}, {Bi}, {Chen}, {Chen}, {Chen}, {Chen}, {Chen}, {Cirennima}, {Cui}, {Danzengluobu}, {Ding}, {Fang}, {Fang}, {Feng}, {Feng}, {Feng}, {Gao}, {Gou}, {Guo}, {He}, {He}, {Hibino}, {Hotta}, {Hu}, {Hu}, {Huang}, {Jia}, {Jiang}, {Jin}, {Kajino}, {Kasahara}, {Katayose}, {Kato}, {Kato}, {Kawata}, {Kozai}, {Labaciren}, {Le}, {Li}, {Li}, {Li}, {Lin}, {Liu}, {Liu}, {Liu}, {Liu}, {Lou}, {Lu}, {Meng}, {Mitsui}, {Munakata}, {Nakamura}, {Nanjo}, {Nishizawa}, {Ohnishi}, {Ohta}, {Ozawa}, {Qian}, {Qu}, {Saito}, {Sakata}, {Sako}, {Sengoku}, {Shao}, {Shibata}, {Shiomi}, {Sugimoto}, {Takita}, {Tan}, {Tateyama}, {Torii}, {Tsuchiya}, {Udo}, {Wang}, {Wu}, {Xue}, {Yagisawa}, {Yamamoto}, {Yang}, {Yuan}, {Zhai}, {Zhang}, {Zhang}, {Zhang}, {Zhang}, {Zhang}, {Zhang}, {Zhang}, {Zhaxisangzhu}, {Zhou}, \& {Tibet AS {\ensuremath{\gamma}} Collaboration}}]{Amenomori2019PhRvL.123e1101A}
{Amenomori}, M., {Bao}, Y.~W., {Bi}, X.~J., {et~al.} 2019, \prl, 123, 051101, \dodoi{10.1103/PhysRevLett.123.051101}

\bibitem[{{Arakawa} {et~al.}(2020){Arakawa}, {Hayashida}, {Khangulyan}, \& {Uchiyama}}]{Arakawa2020ApJ...897...33A}
{Arakawa}, M., {Hayashida}, M., {Khangulyan}, D., \& {Uchiyama}, Y. 2020, \apj, 897, 33, \dodoi{10.3847/1538-4357/ab9368}

\bibitem[{{Atoyan}(1999)}]{Atoyan1999A&A...346L..49A}
{Atoyan}, A.~M. 1999, \aap, 346, L49, \dodoi{10.48550/arXiv.astro-ph/9905204}

\bibitem[{{Atoyan} \& {Aharonian}(1996)}]{Atoyan1996MNRAS.278..525A}
{Atoyan}, A.~M., \& {Aharonian}, F.~A. 1996, \mnras, 278, 525, \dodoi{10.1093/mnras/278.2.525}

\bibitem[{{Atoyan} \& {Nahapetian}(1989)}]{Atoyan1989A&A...219...53A}
{Atoyan}, A.~M., \& {Nahapetian}, A. 1989, \aap, 219, 53

\bibitem[{{Avrorin} {et~al.}(2011){Avrorin}, {Aynutdinov}, {Belolaptikov}, {Berezhnev}, {Bogorodsky}, {Budnev}, {Danilchenko}, {Domogatsky}, {Doroshenko}, {Dyachok}, {Dzhilkibaev}, {Ermakov}, {Fialkovsky}, {Gaponenko}, {Golubkov}, {Gres'}, {Gres'}, {Grishin}, {Grishin}, {Klabukov}, {Klimov}, {Kochanov}, {Konischev}, {Korobchenko}, {Koshechkin}, {Kulepov}, {Kuleshov}, {Kuzmichev}, {Lyashuk}, {Middell}, {Mikheyev}, {Milenin}, {Mirgazov}, {Osipova}, {Pan'kov}, {Pan'kov}, {Panfilov}, {Perevalov}, {Petukhov}, {Pliskovsky}, {Poleschuk}, {Popova}, {Prosin}, {Rozanov}, {Rubtzov}, {Ryabov}, {Sheifler}, {Shirokov}, {Shoibonov}, {Spiering}, {Suvorova}, {Tarashansky}, {Wischnewski}, {Zagorodnikov}, {Zhukov}, \& {Yagunov}}]{Avrorin2011NIMPA.639...30A}
{Avrorin}, A., {Aynutdinov}, V., {Belolaptikov}, I., {et~al.} 2011, Nuclear Instruments and Methods in Physics Research A, 639, 30, \dodoi{10.1016/j.nima.2010.09.137}

\bibitem[{{Bandiera} {et~al.}(2020){Bandiera}, {Bucciantini}, {Mart{\'\i}n}, {Olmi}, \& {Torres}}]{Bandiera2020MNRAS.499.2051B}
{Bandiera}, R., {Bucciantini}, N., {Mart{\'\i}n}, J., {Olmi}, B., \& {Torres}, D.~F. 2020, \mnras, 499, 2051, \dodoi{10.1093/mnras/staa2956}

\bibitem[{{Bandiera} {et~al.}(2023{\natexlab{a}}){Bandiera}, {Bucciantini}, {Mart{\'\i}n}, {Olmi}, \& {Torres}}]{Bandiera2023MNRAS.520.2451B}
---. 2023{\natexlab{a}}, \mnras, 520, 2451, \dodoi{10.1093/mnras/stad134}

\bibitem[{{Bandiera} {et~al.}(2023{\natexlab{b}}){Bandiera}, {Bucciantini}, {Olmi}, \& {Torres}}]{Bandiera2023MNRAS.525.2839B}
{Bandiera}, R., {Bucciantini}, N., {Olmi}, B., \& {Torres}, D.~F. 2023{\natexlab{b}}, \mnras, 525, 2839, \dodoi{10.1093/mnras/stad2387}

\bibitem[{{Bandiera} {et~al.}(2002){Bandiera}, {Neri}, \& {Cesaroni}}]{Bandiera2002A&A...386.1044B}
{Bandiera}, R., {Neri}, R., \& {Cesaroni}, R. 2002, \aap, 386, 1044, \dodoi{10.1051/0004-6361:20020325}

\bibitem[{{Bandiera} {et~al.}(1983){Bandiera}, {Pacini}, \& {Salvati}}]{Bandiera1983A&A...126....7B}
{Bandiera}, R., {Pacini}, F., \& {Salvati}, M. 1983, \aap, 126, 7

\bibitem[{{Bednarek}(2003)}]{Bednarek2003A&A...407....1B}
{Bednarek}, W. 2003, \aap, 407, 1, \dodoi{10.1051/0004-6361:20030929}

\bibitem[{{Bednarek} \& {Bartosik}(2003)}]{Bednarek2003A&A...405..689B}
{Bednarek}, W., \& {Bartosik}, M. 2003, \aap, 405, 689, \dodoi{10.1051/0004-6361:20030593}

\bibitem[{{Bednarek} \& {Protheroe}(1997)}]{Bednarek1997PhRvL..79.2616B}
{Bednarek}, W., \& {Protheroe}, R.~J. 1997, \prl, 79, 2616, \dodoi{10.1103/PhysRevLett.79.2616}

\bibitem[{{Bouyahiaoui} {et~al.}(2020){Bouyahiaoui}, {Kachelrie{\ss}}, \& {Semikoz}}]{Bouyahiaoui2020PhRvD.101l3023B}
{Bouyahiaoui}, M., {Kachelrie{\ss}}, M., \& {Semikoz}, D.~V. 2020, \prd, 101, 123023, \dodoi{10.1103/PhysRevD.101.123023}

\bibitem[{{Bucciantini} {et~al.}(2004){Bucciantini}, {Amato}, {Bandiera}, {Blondin}, \& {Del Zanna}}]{Bucciantini2004A&A...423..253B}
{Bucciantini}, N., {Amato}, E., {Bandiera}, R., {Blondin}, J.~M., \& {Del Zanna}, L. 2004, \aap, 423, 253, \dodoi{10.1051/0004-6361:20040360}

\bibitem[{{Bucciantini} {et~al.}(2011){Bucciantini}, {Arons}, \& {Amato}}]{Bucciantini2011MNRAS.410..381B}
{Bucciantini}, N., {Arons}, J., \& {Amato}, E. 2011, \mnras, 410, 381, \dodoi{10.1111/j.1365-2966.2010.17449.x}

\bibitem[{{Bucciantini} {et~al.}(2023){Bucciantini}, {Ferrazzoli}, {Bachetti}, {Rankin}, {Di Lalla}, {Sgr{\`o}}, {Omodei}, {Kitaguchi}, {Mizuno}, {Gunji}, {Watanabe}, {Baldini}, {Slane}, {Weisskopf}, {Romani}, {Possenti}, {Marshall}, {Silvestri}, {Pacciani}, {Negro}, {Muleri}, {de O{\~n}a Wilhelmi}, {Xie}, {Heyl}, {Pesce-Rollins}, {Wong}, {Pilia}, {Agudo}, {Antonelli}, {Baumgartner}, {Bellazzini}, {Bianchi}, {Bongiorno}, {Bonino}, {Brez}, {Capitanio}, {Castellano}, {Cavazzuti}, {Chen}, {Ciprini}, {Costa}, {De Rosa}, {Del Monte}, {Di Gesu}, {Di Marco}, {Donnarumma}, {Doroshenko}, {Dov{\v{c}}iak}, {Ehlert}, {Enoto}, {Evangelista}, {Fabiani}, {Garcia}, {Hayashida}, {Iwakiri}, {Jorstad}, {Kaaret}, {Karas}, {Kislat}, {Kolodziejczak}, {Krawczynski}, {La Monaca}, {Latronico}, {Liodakis}, {Maldera}, {Manfreda}, {Marin}, {Marinucci}, {Marscher}, {Massaro}, {Matt}, {Mitsuishi}, {Ng}, {O'Dell}, {Oppedisano}, {Papitto}, {Pavlov}, {Peirson}, {Perri}, {Petrucci}, {Poutanen}, {Puccetti}, {Ramsey}, {Ratheesh}, {Roberts}, {Soffitta}, {Spandre}, {Swartz}, {Tamagawa}, {Tavecchio}, {Taverna}, {Tawara}, {Tennant}, {Thomas}, {Tombesi}, {Trois}, {Tsygankov}, {Turolla}, {Vink}, {Wu}, \& {Zane}}]{Bucciantini2023NatAs...7..602B}
{Bucciantini}, N., {Ferrazzoli}, R., {Bachetti}, M., {et~al.} 2023, Nature Astronomy, 7, 602, \dodoi{10.1038/s41550-023-01936-8}

\bibitem[{{Cao} {et~al.}(2021){Cao}, {Aharonian}, {An}, {Axikegu}, {Bai}, {Bai}, {Bao}, {Bastieri}, {Bi}, {Bi}, {Cai}, {Cai}, {Cao}, {Chang}, {Chang}, {Chen}, {Chen}, {Chen}, {Chen}, {Chen}, {Chen}, {Chen}, {Chen}, {Chen}, {Chen}, {Chen}, {Chen}, {Chen}, {Chen}, {Cheng}, {Cheng}, {Cui}, {Cui}, {Cui}, {D'Ettorre Piazzoli}, {Dai}, {Dai}, {Dai}, {Danzengluobu}, {Della Volpe}, {Dong}, {Duan}, {Fan}, {Fan}, {Fan}, {Fang}, {Fang}, {Feng}, {Feng}, {Feng}, {Feng}, {Gao}, {Gao}, {Gao}, {Gao}, {Gao}, {Ge}, {Geng}, {Gong}, {Gou}, {Gu}, {Guo}, {Guo}, {Guo}, {Guo}, {Guo}, {Han}, {He}, {He}, {He}, {He}, {He}, {He}, {Heller}, {Hor}, {Hou}, {Hou}, {Hu}, {Hu}, {Hu}, {Hu}, {Huang}, {Huang}, {Huang}, {Huang}, {Huang}, {Huang}, {Ji}, {Ji}, {Jia}, {Jiang}, {Jiang}, {Jin}, {Ke}, {Kuleshov}, {Levochkin}, {Li}, {Li}, {Li}, {Li}, {Li}, {Li}, {Li}, {Li}, {Li}, {Li}, {Li}, {Li}, {Li}, {Li}, {Li}, {Li}, {Li}, {Li}, {Liang}, {Liang}, {Lin}, {Liu}, {Liu}, {Liu}, {Liu}, {Liu}, {Liu}, {Liu}, {Liu}, {Liu}, {Liu}, {Liu}, {Liu}, {Liu}, {Liu}, {Liu}, {Liu}, {Long}, {Lu}, {Lv}, {Ma}, {Ma}, {Ma}, {Mao}, {Masood}, {Min}, {Mitthumsiri}, {Montaruli}, {Nan}, {Pang}, {Pattarakijwanich}, {Pei}, {Qi}, {Qi}, {Qiao}, {Qin}, {Ruffolo}, {Rulev}, {Saiz}, {Shao}, {Shchegolev}, {Sheng}, {Shi}, {Song}, {Stenkin}, {Stepanov}, {Su}, {Sun}, {Sun}, {Sun}, {Tam}, {Tang}, {Tian}, {Wang}, {Wang}, {Wang}, {Wang}, {Wang}, {Wang}, {Wang}, {Wang}, {Wang}, {Wang}, {Wang}, {Wang}, {Wang}, {Wang}, {Wang}, {Wang}, {Wang}, {Wang}, {Wang}, {Wang}, {Wang}, {Wang}, {Wei}, {Wei}, {Wei}, {Wen}, {Wu}, {Wu}, {Wu}, {Wu}, {Wu}, {Xi}, {Xia}, {Xia}, {Xiang}, {Xiao}, {Xiao}, {Xiao}, {Xin}, {Xin}, {Xing}, {Xu}, {Xu}, {Xue}, {Yan}, {Yan}, {Yang}, {Yang}, {Yang}, {Yang}, {Yang}, {Yang}, {Yang}, {Yao}, {Yao}, {Ye}, {Yin}, {Yin}, {You}, {You}, {Yu}, {Yuan}, {Zeng}, {Zeng}, {Zeng}, {Zeng}, {Zha}, {Zhai}, {Zhang}, {Zhang}, {Zhang}, {Zhang}, {Zhang}, {Zhang}, {Zhang}, {Zhang}, {Zhang}, {Zhang}, {Zhang}, {Zhang}, {Zhang}, {Zhang}, {Zhang}, {Zhang}, {Zhang}, {Zhang}, {Zhang}, {Zhao}, {Zhao}, {Zhao}, {Zhao}, {Zhao}, {Zheng}, {Zheng}, {Zhou}, {Zhou}, {Zhou}, {Zhou}, {Zhou}, {Zhou}, {Zhu}, {Zhu}, {Zhu}, {Zhu}, \& {Zuo}}]{Cao2021Sci...373..425L}
{Cao}, Z., {Aharonian}, F., {An}, Q., {et~al.} 2021, Science, 373, 425, \dodoi{10.1126/science.abg5137}

\bibitem[{{Chen} {et~al.}(2019){Chen}, {Huang}, {Hou}, {Tian}, {Li}, {Yuan}, {Wang}, {Wang}, {Tian}, \& {Liu}}]{Chen2019MNRAS.487.1400C}
{Chen}, B.~Q., {Huang}, Y., {Hou}, L.~G., {et~al.} 2019, \mnras, 487, 1400, \dodoi{10.1093/mnras/stz1357}

\bibitem[{{Chen} {et~al.}(2024{\natexlab{a}}){Chen}, {Liang}, {Liu}, \& {Wang}}]{Chen2024ApJ...976..172C}
{Chen}, X.-B., {Liang}, X.-H., {Liu}, R.-Y., \& {Wang}, X.-Y. 2024{\natexlab{a}}, \apj, 976, 172, \dodoi{10.3847/1538-4357/ad87d2}

\bibitem[{{Chen} {et~al.}(2024{\natexlab{b}}){Chen}, {Liu}, {Wang}, \& {Chang}}]{Chen2024MNRAS.527.7915C}
{Chen}, X.-B., {Liu}, R.-Y., {Wang}, X.-Y., \& {Chang}, X.-C. 2024{\natexlab{b}}, \mnras, 527, 7915, \dodoi{10.1093/mnras/stad3733}

\bibitem[{{Cheng} {et~al.}(1990){Cheng}, {Cheung}, {Lau}, {Yu}, \& {Kwok}}]{Cheng1990JPhG...16.1115C}
{Cheng}, K.~S., {Cheung}, T., {Lau}, M.~M., {Yu}, K.~N., \& {Kwok}, P.~W. 1990, Journal of Physics G Nuclear Physics, 16, 1115, \dodoi{10.1088/0954-3899/16/7/022}

\bibitem[{{Chevalier} \& {Gull}(1975)}]{Chevalier1975ApJ...200..399C}
{Chevalier}, R.~A., \& {Gull}, T.~R. 1975, \apj, 200, 399, \dodoi{10.1086/153802}

\bibitem[{{Cristofari} {et~al.}(2017){Cristofari}, {Gabici}, {Humensky}, {Santander}, {Terrier}, {Parizot}, \& {Casanova}}]{Cristofari2017MNRAS.471..201C}
{Cristofari}, P., {Gabici}, S., {Humensky}, T.~B., {et~al.} 2017, \mnras, 471, 201, \dodoi{10.1093/mnras/stx1574}

\bibitem[{{Del Zanna} {et~al.}(2004){Del Zanna}, {Amato}, \& {Bucciantini}}]{DelZanna2004A&A...421.1063D}
{Del Zanna}, L., {Amato}, E., \& {Bucciantini}, N. 2004, \aap, 421, 1063, \dodoi{10.1051/0004-6361:20035936}

\bibitem[{{Di Palma} {et~al.}(2017){Di Palma}, {Guetta}, \& {Amato}}]{DiPalma2017ApJ...836..159D}
{Di Palma}, I., {Guetta}, D., \& {Amato}, E. 2017, \apj, 836, 159, \dodoi{10.3847/1538-4357/836/2/159}

\bibitem[{{Dirson} \& {Horns}(2023)}]{Dirson2023A&A...671A..67D}
{Dirson}, L., \& {Horns}, D. 2023, \aap, 671, A67, \dodoi{10.1051/0004-6361/202243578}

\bibitem[{{Drury}(1983)}]{Drury1983RPPh...46..973D}
{Drury}, L.~O. 1983, Reports on Progress in Physics, 46, 973, \dodoi{10.1088/0034-4885/46/8/002}

\bibitem[{{Evoli} {et~al.}(2007){Evoli}, {Grasso}, \& {Maccione}}]{Evoli2007JCAP...06..003E}
{Evoli}, C., {Grasso}, D., \& {Maccione}, L. 2007, \jcap, 2007, 003, \dodoi{10.1088/1475-7516/2007/06/003}

\bibitem[{{Faucher-Gigu{\`e}re} \& {Kaspi}(2006)}]{Faucher-Giguere2006ApJ...643..332F}
{Faucher-Gigu{\`e}re}, C.-A., \& {Kaspi}, V.~M. 2006, \apj, 643, 332, \dodoi{10.1086/501516}

\bibitem[{{Fiori} {et~al.}(2022){Fiori}, {Olmi}, {Amato}, {Bandiera}, {Bucciantini}, {Zampieri}, \& {Burtovoi}}]{Fiori2022MNRAS.511.1439F}
{Fiori}, M., {Olmi}, B., {Amato}, E., {et~al.} 2022, \mnras, 511, 1439, \dodoi{10.1093/mnras/stac019}

\bibitem[{{Fouka} \& {Ouichaoui}(2013)}]{Fouka2013RAA....13..680F}
{Fouka}, M., \& {Ouichaoui}, S. 2013, Research in Astronomy and Astrophysics, 13, 680, \dodoi{10.1088/1674-4527/13/6/007}

\bibitem[{{Gaensler} \& {Slane}(2006)}]{Gaensler2006ARA&A..44...17G}
{Gaensler}, B.~M., \& {Slane}, P.~O. 2006, \araa, 44, 17, \dodoi{10.1146/annurev.astro.44.051905.092528}

\bibitem[{{Gaggero} {et~al.}(2015){Gaggero}, {Grasso}, {Marinelli}, {Urbano}, \& {Valli}}]{Gaggero2015ApJ...815L..25G}
{Gaggero}, D., {Grasso}, D., {Marinelli}, A., {Urbano}, A., \& {Valli}, M. 2015, \apjl, 815, L25, \dodoi{10.1088/2041-8205/815/2/L25}

\bibitem[{{Gagliardini} {et~al.}(2024){Gagliardini}, {Langella}, {Guetta}, \& {Capone}}]{Gagliardini2024ApJ...969..161G}
{Gagliardini}, S., {Langella}, A., {Guetta}, D., \& {Capone}, A. 2024, \apj, 969, 161, \dodoi{10.3847/1538-4357/ad4960}

\bibitem[{{Gelfand} {et~al.}(2009){Gelfand}, {Slane}, \& {Zhang}}]{Gelfand2009ApJ...703.2051G}
{Gelfand}, J.~D., {Slane}, P.~O., \& {Zhang}, W. 2009, \apj, 703, 2051, \dodoi{10.1088/0004-637X/703/2/2051}

\bibitem[{{Giacinti} \& {Semikoz}(2023)}]{Giacinti2023arXiv230510251G}
{Giacinti}, G., \& {Semikoz}, D. 2023, arXiv e-prints, arXiv:2305.10251, \dodoi{10.48550/arXiv.2305.10251}

\bibitem[{{Guetta} \& {Amato}(2003)}]{Guetta2003APh....19..403G}
{Guetta}, D., \& {Amato}, E. 2003, Astroparticle Physics, 19, 403, \dodoi{10.1016/S0927-6505(02)00221-9}

\bibitem[{{Hester}(2008)}]{Hester2008ARA&A..46..127H}
{Hester}, J.~J. 2008, \araa, 46, 127, \dodoi{10.1146/annurev.astro.45.051806.110608}

\bibitem[{{Hester} {et~al.}(1996){Hester}, {Stone}, {Scowen}, {Jun}, {Gallagher}, {Norman}, {Ballester}, {Burrows}, {Casertano}, {Clarke}, {Crisp}, {Griffiths}, {Hoessel}, {Holtzman}, {Krist}, {Mould}, {Sankrit}, {Stapelfeldt}, {Trauger}, {Watson}, \& {Westphal}}]{Hester1996ApJ...456..225H}
{Hester}, J.~J., {Stone}, J.~M., {Scowen}, P.~A., {et~al.} 1996, \apj, 456, 225, \dodoi{10.1086/176643}

\bibitem[{{Hinton} {et~al.}(2011){Hinton}, {Funk}, {Parsons}, \& {Ohm}}]{Hinton2011ApJ...743L...7H}
{Hinton}, J.~A., {Funk}, S., {Parsons}, R.~D., \& {Ohm}, S. 2011, \apjl, 743, L7, \dodoi{10.1088/2041-8205/743/1/L7}

\bibitem[{{Horns} {et~al.}(2006){Horns}, {Aharonian}, {Santangelo}, {Hoffmann}, \& {Masterson}}]{Horns2006A&A...451L..51H}
{Horns}, D., {Aharonian}, F., {Santangelo}, A., {Hoffmann}, A.~I.~D., \& {Masterson}, C. 2006, \aap, 451, L51, \dodoi{10.1051/0004-6361:20065116}

\bibitem[{{Hou} \& {Han}(2014)}]{Hou2014A&A...569A.125H}
{Hou}, L.~G., \& {Han}, J.~L. 2014, \aap, 569, A125, \dodoi{10.1051/0004-6361/201424039}

\bibitem[{{Huang} {et~al.}(2024){Huang}, {Cao}, {Chen}, {Liu}, {Wang}, {You}, \& {Qi}}]{Huang2024icrc.confE1080H}
{Huang}, T.~Q., {Cao}, Z., {Chen}, M., {et~al.} 2024, in 38th International Cosmic Ray Conference, 1080

\bibitem[{{Johnston} {et~al.}(2020){Johnston}, {Smith}, {Karastergiou}, \& {Kramer}}]{Johnston2020MNRAS.497.1957J}
{Johnston}, S., {Smith}, D.~A., {Karastergiou}, A., \& {Kramer}, M. 2020, \mnras, 497, 1957, \dodoi{10.1093/mnras/staa2110}

\bibitem[{{Jones}(1968)}]{Jones1968PhRv..167.1159J}
{Jones}, F.~C. 1968, Physical Review, 167, 1159, \dodoi{10.1103/PhysRev.167.1159}

\bibitem[{{Jun}(1998)}]{Jun1998ApJ...499..282J}
{Jun}, B.-I. 1998, \apj, 499, 282, \dodoi{10.1086/305627}

\bibitem[{{Kasen} \& {Woosley}(2009)}]{Kasen2009ApJ...703.2205K}
{Kasen}, D., \& {Woosley}, S.~E. 2009, \apj, 703, 2205, \dodoi{10.1088/0004-637X/703/2/2205}

\bibitem[{{Kennel} \& {Coroniti}(1984{\natexlab{a}})}]{Kennel1984ApJ...283..694K}
{Kennel}, C.~F., \& {Coroniti}, F.~V. 1984{\natexlab{a}}, \apj, 283, 694, \dodoi{10.1086/162356}

\bibitem[{{Kennel} \& {Coroniti}(1984{\natexlab{b}})}]{Kennel1984ApJ...283..710K}
---. 1984{\natexlab{b}}, \apj, 283, 710, \dodoi{10.1086/162357}

\bibitem[{{Koldobskiy} {et~al.}(2021){Koldobskiy}, {Kachelrie{\ss}}, {Lskavyan}, {Neronov}, {Ostapchenko}, \& {Semikoz}}]{Koldobskiy2021PhRvD.104l3027K}
{Koldobskiy}, S., {Kachelrie{\ss}}, M., {Lskavyan}, A., {et~al.} 2021, \prd, 104, 123027, \dodoi{10.1103/PhysRevD.104.123027}

\bibitem[{{Kovalev} {et~al.}(2022){Kovalev}, {Plavin}, \& {Troitsky}}]{Kovalev2022ApJ...940L..41K}
{Kovalev}, Y.~Y., {Plavin}, A.~V., \& {Troitsky}, S.~V. 2022, \apjl, 940, L41, \dodoi{10.3847/2041-8213/aca1ae}

\bibitem[{{LHAASO Collaboration}(2024)}]{LHAASO2024SciBu..69..449L}
{LHAASO Collaboration}. 2024, Science Bulletin, 69, 449, \dodoi{10.1016/j.scib.2023.12.040}

\bibitem[{{Liang} {et~al.}(2022){Liang}, {Li}, {Wu}, {Pan}, \& {Liu}}]{Liang2022Univ....8..547L}
{Liang}, X.-H., {Li}, C.-M., {Wu}, Q.-Z., {Pan}, J.-S., \& {Liu}, R.-Y. 2022, Universe, 8, 547, \dodoi{10.3390/universe8100547}

\bibitem[{{Lipari} \& {Vernetto}(2018)}]{Lipari2018PhRvD..98d3003L}
{Lipari}, P., \& {Vernetto}, S. 2018, \prd, 98, 043003, \dodoi{10.1103/PhysRevD.98.043003}

\bibitem[{{Liu} \& {Wang}(2021)}]{Liu2021ApJ...922..221L}
{Liu}, R.-Y., \& {Wang}, X.-Y. 2021, \apj, 922, 221, \dodoi{10.3847/1538-4357/ac2ba0}

\bibitem[{{Lu} {et~al.}(2017){Lu}, {Gao}, \& {Zhang}}]{Lu2017ApJ...834...43L}
{Lu}, F.-W., {Gao}, Q.-G., \& {Zhang}, L. 2017, \apj, 834, 43, \dodoi{10.3847/1538-4357/834/1/43}

\bibitem[{{Luo} {et~al.}(2020){Luo}, {Lyutikov}, {Temim}, \& {Comisso}}]{Luo2020ApJ...896..147L}
{Luo}, Y., {Lyutikov}, M., {Temim}, T., \& {Comisso}, L. 2020, \apj, 896, 147, \dodoi{10.3847/1538-4357/ab93c0}

\bibitem[{{Lyutikov} {et~al.}(2019){Lyutikov}, {Temim}, {Komissarov}, {Slane}, {Sironi}, \& {Comisso}}]{Lyutikov2019MNRAS.489.2403L}
{Lyutikov}, M., {Temim}, T., {Komissarov}, S., {et~al.} 2019, \mnras, 489, 2403, \dodoi{10.1093/mnras/stz2023}

\bibitem[{{Martin} \& {Torres}(2022)}]{Martin2022JHEAp..36..128M}
{Martin}, J., \& {Torres}, D.~F. 2022, Journal of High Energy Astrophysics, 36, 128, \dodoi{10.1016/j.jheap.2022.09.003}

\bibitem[{{Mart{\'\i}n} {et~al.}(2012){Mart{\'\i}n}, {Torres}, \& {Rea}}]{Martin2012MNRAS.427..415M}
{Mart{\'\i}n}, J., {Torres}, D.~F., \& {Rea}, N. 2012, \mnras, 427, 415, \dodoi{10.1111/j.1365-2966.2012.22014.x}

\bibitem[{{Martin} {et~al.}(2022){Martin}, {Tibaldo}, {Marcowith}, \& {Abdollahi}}]{Martin2022A&A...666A...7M}
{Martin}, P., {Tibaldo}, L., {Marcowith}, A., \& {Abdollahi}, S. 2022, \aap, 666, A7, \dodoi{10.1051/0004-6361/202244002}

\bibitem[{{Martin} {et~al.}(2021){Martin}, {Milisavljevic}, \& {Drissen}}]{Martin2021MNRAS.502.1864M}
{Martin}, T., {Milisavljevic}, D., \& {Drissen}, L. 2021, \mnras, 502, 1864, \dodoi{10.1093/mnras/staa4046}

\bibitem[{{Meagher}(2015)}]{Meagher2015ICRC...34..792M}
{Meagher}, K. 2015, in International Cosmic Ray Conference, Vol.~34, 34th International Cosmic Ray Conference (ICRC2015), 792, \dodoi{10.22323/1.236.0792}

\bibitem[{{Meyer} {et~al.}(2010){Meyer}, {Horns}, \& {Zechlin}}]{Meyer2010A&A...523A...2M}
{Meyer}, M., {Horns}, D., \& {Zechlin}, H.~S. 2010, \aap, 523, A2, \dodoi{10.1051/0004-6361/201014108}

\bibitem[{{Neronov} {et~al.}(2023){Neronov}, {Semikoz}, {Aublin}, {Lamoureux}, \& {Kouchner}}]{Neronov2023PhRvD.108j3044N}
{Neronov}, A., {Semikoz}, D., {Aublin}, J., {Lamoureux}, M., \& {Kouchner}, A. 2023, \prd, 108, 103044, \dodoi{10.1103/PhysRevD.108.103044}

\bibitem[{{Nie} {et~al.}(2022){Nie}, {Liu}, {Jiang}, \& {Geng}}]{Nie2022ApJ...924...42N}
{Nie}, L., {Liu}, Y., {Jiang}, Z., \& {Geng}, X. 2022, \apj, 924, 42, \dodoi{10.3847/1538-4357/ac348d}

\bibitem[{{Nodes} {et~al.}(2004){Nodes}, {Birk}, {Gritschneder}, \& {Lesch}}]{Nodes2004A&A...423...13N}
{Nodes}, C., {Birk}, G.~T., {Gritschneder}, M., \& {Lesch}, H. 2004, \aap, 423, 13, \dodoi{10.1051/0004-6361:20047065}

\bibitem[{{Olmi} {et~al.}(2014){Olmi}, {Del Zanna}, {Amato}, {Bandiera}, \& {Bucciantini}}]{Olmi2014MNRAS.438.1518O}
{Olmi}, B., {Del Zanna}, L., {Amato}, E., {Bandiera}, R., \& {Bucciantini}, N. 2014, \mnras, 438, 1518, \dodoi{10.1093/mnras/stt2308}

\bibitem[{{Olmi} {et~al.}(2016){Olmi}, {Del Zanna}, {Amato}, {Bucciantini}, \& {Mignone}}]{Olmi2016JPlPh..82f6301O}
{Olmi}, B., {Del Zanna}, L., {Amato}, E., {Bucciantini}, N., \& {Mignone}, A. 2016, Journal of Plasma Physics, 82, 635820601, \dodoi{10.1017/S0022377816000957}

\bibitem[{{Owen} \& {Barlow}(2015)}]{Owen2015ApJ...801..141O}
{Owen}, P.~J., \& {Barlow}, M.~J. 2015, \apj, 801, 141, \dodoi{10.1088/0004-637X/801/2/141}

\bibitem[{{Parker}(1965)}]{Parker1965P&SS...13....9P}
{Parker}, E.~N. 1965, \planss, 13, 9, \dodoi{10.1016/0032-0633(65)90131-5}

\bibitem[{{Peng} {et~al.}(2022){Peng}, {Bao}, {Lu}, \& {Zhang}}]{Peng2022ApJ...926....7P}
{Peng}, Q.-Y., {Bao}, B.-W., {Lu}, F.-W., \& {Zhang}, L. 2022, \apj, 926, 7, \dodoi{10.3847/1538-4357/ac4161}

\bibitem[{{Porth} {et~al.}(2014{\natexlab{a}}){Porth}, {Komissarov}, \& {Keppens}}]{Porth2014MNRAS.438..278P}
{Porth}, O., {Komissarov}, S.~S., \& {Keppens}, R. 2014{\natexlab{a}}, \mnras, 438, 278, \dodoi{10.1093/mnras/stt2176}

\bibitem[{{Porth} {et~al.}(2014{\natexlab{b}}){Porth}, {Komissarov}, \& {Keppens}}]{Porth2014MNRAS.443..547P}
---. 2014{\natexlab{b}}, \mnras, 443, 547, \dodoi{10.1093/mnras/stu1082}

\bibitem[{{Reid} {et~al.}(2019){Reid}, {Menten}, {Brunthaler}, {Zheng}, {Dame}, {Xu}, {Li}, {Sakai}, {Wu}, {Immer}, {Zhang}, {Sanna}, {Moscadelli}, {Rygl}, {Bartkiewicz}, {Hu}, {Quiroga-Nu{\~n}ez}, \& {van Langevelde}}]{Reid2019ApJ...885..131R}
{Reid}, M.~J., {Menten}, K.~M., {Brunthaler}, A., {et~al.} 2019, \apj, 885, 131, \dodoi{10.3847/1538-4357/ab4a11}

\bibitem[{{Smartt}(2009)}]{Smartt2009ARA&A..47...63S}
{Smartt}, S.~J. 2009, \araa, 47, 63, \dodoi{10.1146/annurev-astro-082708-101737}

\bibitem[{{Smith}(2013)}]{Smith2013MNRAS.434..102S}
{Smith}, N. 2013, \mnras, 434, 102, \dodoi{10.1093/mnras/stt1004}

\bibitem[{{Stecker}(1979)}]{Stecker1979ApJ...228..919S}
{Stecker}, F.~W. 1979, \apj, 228, 919, \dodoi{10.1086/156919}

\bibitem[{{Stockinger} {et~al.}(2020){Stockinger}, {Janka}, {Kresse}, {Melson}, {Ertl}, {Gabler}, {Gessner}, {Wongwathanarat}, {Tolstov}, {Leung}, {Nomoto}, \& {Heger}}]{Stockinger2020MNRAS.496.2039S}
{Stockinger}, G., {Janka}, H.~T., {Kresse}, D., {et~al.} 2020, \mnras, 496, 2039, \dodoi{10.1093/mnras/staa1691}

\bibitem[{{Tanaka} \& {Asano}(2017)}]{Tanaka2017ApJ...841...78T}
{Tanaka}, S.~J., \& {Asano}, K. 2017, \apj, 841, 78, \dodoi{10.3847/1538-4357/aa6f13}

\bibitem[{{Tanaka} \& {Takahara}(2010)}]{Tanaka2010ApJ...715.1248T}
{Tanaka}, S.~J., \& {Takahara}, F. 2010, \apj, 715, 1248, \dodoi{10.1088/0004-637X/715/2/1248}

\bibitem[{{Tauris} \& {Manchester}(1998)}]{Tauris1998MNRAS.298..625T}
{Tauris}, T.~M., \& {Manchester}, R.~N. 1998, \mnras, 298, 625, \dodoi{10.1046/j.1365-8711.1998.01369.x}

\bibitem[{{Temim} {et~al.}(2024){Temim}, {Laming}, {Kavanagh}, {Smith}, {Slane}, {Blair}, {De Looze}, {Bucciantini}, {Jerkstrand}, {Gountanis}, {Sankrit}, {Milisavljevic}, {Rest}, {Lyutikov}, {DePasquale}, {Martin}, {Drissen}, {Raymond}, {Fox}, {Modjaz}, {Spitkovsky}, \& {Strolger}}]{Temim2024ApJ...968L..18T}
{Temim}, T., {Laming}, J.~M., {Kavanagh}, P.~J., {et~al.} 2024, \apjl, 968, L18, \dodoi{10.3847/2041-8213/ad50d1}

\bibitem[{{Torres} {et~al.}(2014){Torres}, {Cillis}, {Mart{\'\i}n}, \& {de O{\~n}a Wilhelmi}}]{Torres2014JHEAp...1...31T}
{Torres}, D.~F., {Cillis}, A., {Mart{\'\i}n}, J., \& {de O{\~n}a Wilhelmi}, E. 2014, Journal of High Energy Astrophysics, 1, 31, \dodoi{10.1016/j.jheap.2014.02.001}

\bibitem[{{Torres} {et~al.}(2013){Torres}, {Mart{\'\i}n}, {de O{\~n}a Wilhelmi}, \& {Cillis}}]{Torres2013MNRAS.436.3112T}
{Torres}, D.~F., {Mart{\'\i}n}, J., {de O{\~n}a Wilhelmi}, E., \& {Cillis}, A. 2013, \mnras, 436, 3112, \dodoi{10.1093/mnras/stt1793}

\bibitem[{{Troitsky}(2024)}]{Troitsky2024PhyU...67..349T}
{Troitsky}, S.~V. 2024, Physics Uspekhi, 67, 349, \dodoi{10.3367/UFNe.2023.04.039581}

\bibitem[{{Truelove} \& {McKee}(1999)}]{Truelove1999ApJS..120..299T}
{Truelove}, J.~K., \& {McKee}, C.~F. 1999, \apjs, 120, 299, \dodoi{10.1086/313176}

\bibitem[{{Vorster} \& {Moraal}(2013)}]{Vorster2013ApJ...765...30V}
{Vorster}, M.~J., \& {Moraal}, H. 2013, \apj, 765, 30, \dodoi{10.1088/0004-637X/765/1/30}

\bibitem[{{Vorster} {et~al.}(2013){Vorster}, {Tibolla}, {Ferreira}, \& {Kaufmann}}]{Vorster2013ApJ...773..139V}
{Vorster}, M.~J., {Tibolla}, O., {Ferreira}, S.~E.~S., \& {Kaufmann}, S. 2013, \apj, 773, 139, \dodoi{10.1088/0004-637X/773/2/139}

\bibitem[{{Watters} \& {Romani}(2011)}]{Watters2011ApJ...727..123W}
{Watters}, K.~P., \& {Romani}, R.~W. 2011, \apj, 727, 123, \dodoi{10.1088/0004-637X/727/2/123}

\bibitem[{{Weisskopf} {et~al.}(2012){Weisskopf}, {Elsner}, {Kolodziejczak}, {O'Dell}, \& {Tennant}}]{Weisskopf2012ApJ...746...41W}
{Weisskopf}, M.~C., {Elsner}, R.~F., {Kolodziejczak}, J.~J., {O'Dell}, S.~L., \& {Tennant}, A.~F. 2012, \apj, 746, 41, \dodoi{10.1088/0004-637X/746/1/41}

\bibitem[{{Xie} {et~al.}(2024){Xie}, {Wang}, {Wang}, {Manchester}, \& {Hobbs}}]{Xie2024ApJ...963L..39X}
{Xie}, J.~T., {Wang}, J.~B., {Wang}, N., {Manchester}, R., \& {Hobbs}, G. 2024, \apjl, 963, L39, \dodoi{10.3847/2041-8213/ad2850}

\bibitem[{{Yao} {et~al.}(2017){Yao}, {Manchester}, \& {Wang}}]{Yao2017ApJ...835...29Y}
{Yao}, J.~M., {Manchester}, R.~N., \& {Wang}, N. 2017, \apj, 835, 29, \dodoi{10.3847/1538-4357/835/1/29}

\bibitem[{{Ye} {et~al.}(2023){Ye}, {Hu}, {Tian}, {Chang}, {Chang}, {Cheng}, {Gao}, {Ge}, {Gong}, {Guo}, {Guo}, {He}, {Huang}, {Jiang}, {Jiang}, {Jing}, {Li}, {Li}, {Li}, {Li}, {Li}, {Liao}, {Lin}, {Lin}, {Liu}, {Liu}, {Liu}, {Miao}, {Mo}, {Morton-Blake}, {Peng}, {Sun}, {Tang}, {Tang}, {Tao}, {Tian}, {Wang}, {Wang}, {Wang}, {Wei}, {Wei}, {Wu}, {Xian}, {Xiang}, {Xu}, {Xue}, {Yang}, {Yang}, {Yu}, {Zeng}, {Zhang}, {Zhang}, {Zhang}, {Zhang}, {Zhi}, {Zhong}, {Zhou}, {Zhu}, \& {Zhuang}}]{Ye2023NatAs...7.1497Y}
{Ye}, Z.~P., {Hu}, F., {Tian}, W., {et~al.} 2023, Nature Astronomy, 7, 1497, \dodoi{10.1038/s41550-023-02087-6}

\bibitem[{{You} {et~al.}(2024){You}, {Chen}, {Pan}, {Tsai}, \& {Ou}}]{You2024ApJ...970..145Y}
{You}, K.-A., {Chen}, K.-J., {Pan}, Y.-C., {Tsai}, S.-H., \& {Ou}, P.-S. 2024, \apj, 970, 145, \dodoi{10.3847/1538-4357/ad50c6}

\bibitem[{{Yusifov} \& {K{\"u}{\c{c}}{\"u}k}(2004)}]{Yusifov2004A&A...422..545Y}
{Yusifov}, I., \& {K{\"u}{\c{c}}{\"u}k}, I. 2004, \aap, 422, 545, \dodoi{10.1051/0004-6361:20040152}

\bibitem[{{Zhang} \& {Yang}(2009)}]{Zhang2009ApJ...699L.153Z}
{Zhang}, L., \& {Yang}, X.~C. 2009, \apjl, 699, L153, \dodoi{10.1088/0004-637X/699/2/L153}

\bibitem[{{Zhang} {et~al.}(2020){Zhang}, {Chen}, {Huang}, \& {Chen}}]{Zhang2020MNRAS.497.3477Z}
{Zhang}, X., {Chen}, Y., {Huang}, J., \& {Chen}, D. 2020, \mnras, 497, 3477, \dodoi{10.1093/mnras/staa2151}

\end{thebibliography}
\bibliographystyle{aasjournal}

\end{document}